\documentclass{aa}  

\usepackage{graphicx}
\usepackage[export]{adjustbox}
\usepackage[dvipsnames]{xcolor}
\usepackage{txfonts}
\bibpunct{(}{)}{;}{a}{}{,}
\newcommand{\ddr}[1]{\frac{\partial{}#1}{\partial{}r}}

\begin{document}

   \title{Co-existence of Internal Gravity Waves and Tayler-Spruit \\ Magnetic Fields in the Radiative Core of Low-mass Stars}

   \titlerunning{IGW in Tayler-Spruit Magnetic Field}

   \subtitle{}

   \author{L. Amard\inst{1,2}
          \and
          S. Mathis\inst{2}
          }

   \institute{Department of Astronomy, University of Geneva, Chemin Pegasi 51, 1290 Versoix, Switzerland 
   \and 
   Universit\'e Paris-Saclay, Universit\'e Paris Cit\'e, CEA, CNRS, AIM, F-91191, Gif-sur-Yvette, France \\
   \email{louis.amard@unige.ch}
}

   \date{Received December 1st, 2024; accepted March 16, 2025}

 
  \abstract
   {The Tayler-Spruit dynamo (TSD) is able to generate a small-scale magnetic field in the differentially rotating stably stratified layers of stars and was recently observed in numerical simulations. In parallel, the propagation of internal gravity waves in stars can be modified in the presence of a magnetic field.}
   {We first want to estimate the interaction between a magnetic field generated by the TSD and internal gravity waves in the radiative core of low-mass stars. This allows us to characterise the effect of this interplay on the observed standing modes spectrum and on the internal transport of angular momentum by progressive waves.}
   {We use the STAREVOL evolution code to compute the structure of low-mass rotating stars with various transport and mixing {mechanisms} along their evolution. We implement the formalism to describe the TSD and estimate the regions in which the magnetic field generated by TSD is strong enough to change the identity of internal gravity waves to magneto-gravity waves. {In addition, we evaluate the possible limitation of angular momentum transport by the combined action of rotation and magnetism.}}
   {Along the pre-main sequence and main-sequence evolution of low-mass stars, the lowest frequencies of the excited gravity wave spectrum should be converted to magneto-gravity waves by the magnetic field generated by the TSD. During the red-giant branch we find that most of the excited spectrum of progressive internal gravity waves could be converted into magneto-gravity waves.}
   {}


   \maketitle
%

\section{Introduction}
The evolution of stellar angular momentum is a fundamental question in astrophysics. The rotation drives a wide variety of physical mechanisms that affects the life of stars \citep{Maederbook}. 
Helio- and now asteroseismology allow us to probe the inner regions of the Sun and other stars. In particular, we can estimate the solar rotation profile {in its radiative core }down to 20\% of its radius \citep[\textit{e.g.}][]{Couvidat2003,Garcia2007,Garcia2011,Korzennik2024}. It was found a rotation profile very close to solid body, indicative of a very efficient redistribution of angular momentum. In addition, thanks to mixed modes being able to reach the surface, we can retrieve the rotation rate of the inner regions of evolved stars \citep{Beck2012,Mosser2012,Deheuvelsetal2014,Deheuvelsetal2015,Gehan2018,Deheuvels2020,Mosser2024,Li2024}. Once again, the shallow differential rotation between the core and the surface defies most angular momentum transport models. 

The transport of angular momentum in radiative region has been largely debated and several physical processes have been identified over the years to try to explain the rotation profile of the Sun and other type of stars \citep{CharbonneauMacGregor1993,GoughMcIntyre1998,CharbonnelTalon2005Science,Eggenberger2005,Strugarek2011,Cantiello2014,Fuller2014,Fuller2019,Belkacem2015,Eggenberger2019a,Eggenberger2019c,Eggenberger2019b,Garaud2024}.  While the transport of angular momentum by purely hydrodynamic processes is not efficient enough to explain the rotation profiles measured in the Sun and red giants thanks to helio- and asteroseismology, {respectively} \citep{TurckChieze2010,Eggenbergeretal12,Cellieretal2012,Marques2013,Cantiello2014,Mathis2018}, other mechanisms have been explored.

On the one hand, stellar interiors are likely magnetised. When they are combined with a mild differential rotation, Maxwell stresses have been shown to efficiently redistribute angular momentum even with low magnetic fields \citep{MestelWeiss1987,Mestel1988}. 
\cite{Spruit2002} then suggested a loop in which the Tayler instability of an initial azimuthal magnetic field \citep{Tayler1973} is self-sustained and generates a magnetic field in the stellar radiation zone, even with a mild differential rotation. The corresponding formalism can be implemented in stellar evolution codes \citep{MaederMeynet2004,Heger2005,Petrovic2005,Eggenberger2005,Eggenberger2019b}. In particular, \citet{Eggenberger2005} showed  that this instability could explain the efficient transport in the solar interior {necessary to} match the rotation profile reconstructed from helioseismology.
\cite{Fuller2019} suggested a revised prescription where the main difference with \cite{Spruit2002} is the saturation for the Tayler instability. They propose a different turbulence cascade that leads to lower energy dissipation rate, and therefore to higher magnetic field amplitude and angular momentum transport.
Unfortunately, the transport associated to both formalisms remains unable to explain simultaneously red giants' and sub-giants' rotation  profiles \citep{Cantiello2014,Eggenberger2019c}. 
Until recently, this Tayler-Spruit dynamo (hereafter TSD) loop was not {observed} in MHD simulations \citep{Zahn2007}. \cite{Petitdemange2023} first {identified} in their simulations a mechanism with very similar properties to the TSD, with a {field} strength comparable to what is predicted by the formalism by \cite{Spruit2002}. A year later, \cite{Barrere2024}  and \cite{Petitdemange2024} explored the instability over a wider range of parameters and {identified} branches of the instability {that scaled} similarly to the prescription by \cite{Fuller2019}.

On the other hand, internal gravity waves {(hereafter IGW)} excited at the interface between convective and radiative layers can propagate in the latter and efficiently carry angular momentum to the location {where} they are damped \citep{Schatzman1993,Zahn1997,KTZ1999,TalonCharbonnel2005,Pincon2017}. {This mechanism can potentially explain the chemical mixing and the rotational profile in the Sun and solar-type stars \citep{CharbonnelTalon2005Science,TalonCharbonnel2010}.}

{Internal magnetic fields and gravity waves} have {thus} both shown the potential to explain the rotation profile of the Sun and low-mass stars. Nevertheless, while both mechanisms are expected to be present, their interplay has never been studied systematically along the evolution of these stars.

In this framework, one of the key {results} of the \textit{Kepler} mission was that about 40\% of studied {red} giants have a missing part in their spectrum \citep[e.g.][]{Garciaetal2014,Stello2016,Bugnet2022}. The magnetic field has been proposed as the main mechanism to explain these so-called 'depressed' stars. Indeed, \cite{Fuller2015} demonstrated that a strong enough magnetic field may trap some of the gravity modes in the inner regions and convert them to Alfv\'en waves, changing their identity and preventing the modes to travel back to the surface, leading to a missing part of the spectrum.
  
Several questions then arise. First, how can we describe the interactions between IGW and a magnetic field generated by TSD? Then, how do they affect the progressive waves and the standing modes along the evolution? And finally, how efficient are the modified waves at transporting angular momentum?

A lot of efforts has been made in the last decade {to study the interaction between waves and a magnetic field. On the theoretical side we }refer to{ the works by \cite{GoughThompson1990,Bugnet2021,Mathis2021,Li2022}, and \cite{Lignieres2024} with a perturbative approach, and by \cite{RogersMacGregor2010,MDB2011,MDB2012,Lecoanet2017,LoiPapaloizou2018,Loi2020a,Loi2020b,Loi2021,Dhouib2022,RuiFuller2023,Rui2024} with a non-perturbative approach.} The conversion mechanism from gravity waves to Alfv\'en waves, \textit{i.e.} the change of identity of the wave is presented in \cite{Fuller2015} and \cite{RuiFuller2023}, and confirmed by the simulations of \cite{Lecoanet2017} \citep[See also the recent work by][]{David2025}, while the impact on the transport of angular momentum can be found in \cite{MDB2012}. In particular, the predicted amplitudes {of} the critical field at which {IGWs} are converted in Magneto-GW have already been used to constrain the magnetic field in the core of red giant {and massive} stars \citep{Cantiello2016,Bugnet2021,Li2022,Lecoanet2022}. 

In order to study the role of the combined mechanism along the evolution, we decided to follow the description of the TSD by \cite{MaederMeynet2004} which has been confirmed recently by the {numerical} simulations of \cite{Petitdemange2023}. 
In this work, we {aim} to understand the various identity changes of the waves along the structure of evolving stars with a radiative core. We want to estimate the role of each mechanism on the transport of angular momentum and the seismic signature, when both {IGW} and a {TSD} co-exist in the radiative zone.
We first describe the type of waves, the magnetic field properties and the physics of their interaction in stellar stably stratified radiative regions. In Section~\S 3 we detail the application to the case of a sun-like 1$M_\odot$ star along its evolution. We then extend the discussion to the case of {solar-type stars} in~\S 4 before concluding. 

\section{Physics of the problem}
\label{Sect:Physics}
Several timescales are to be considered in a magnetised stably stratified rotating radiative region. First, the buoyancy stabilises the medium following the Brunt-V\"ais\"al\"a frequency $N$. Then, unless the star is close to breakup (which we do not consider here), below is the rotation frequency $2\Omega$ with $\Omega$ the rotation rate, and finally, as displayed in figure~\ref{fig:diagfreq}, the Alfv\'en frequency $\omega_A$ {is generally the lowest with} 
\begin{equation}
    \omega_A\equiv\Vec{k}\cdot\vec{V}_A,
    \label{eq:omega_A}
\end{equation}
{where $\vec{k}$ is the wave vector and $\vec{V}_A$ the Alfv\'en velocity defined as $\vec{V}_A=\Vec{B}/\sqrt{4\pi\rho}$, with $\rho$ and $\Vec{B}$ being the density and the local magnetic field in cgs units, respectively.}
 This leads to several possible types of waves and physical mechanisms occurring in this region.
In addition, we assume a weak differential rotation in the latitudinal direction due to anisotropic turbulence and magnetic instabilities \citep{Zahn1992,Spruit2002,Mathis2018}.

\subsection{Internal gravity waves}
{IGW} propagate in stably stratified regions such as the radiative core of solar-type stars with a frequency $\omega<N$. 
{They are stochastically excited by convective Reynold stresses} \citep{Press1981,Belkacem2009,LecoanetQuataert2013} and by the penetration of convective plumes in the radiative layers \citep{Rogers2013,Alvan2014,Pincon2016}. 
The {strength of the} two excitation mechanisms are comparable all along the evolution in terms of energy injected, but they excite different parts of the wave spectrum. In particular, the energy spectrum is relatively narrow and peaks at low $l$-degree in the case of an excitation by turbulent {convective Reynold stresses}, while it is broader and shifted towards higher angular degrees with penetrative convection \citep[see fig. 8 in][]{Pincon2016}.
The wave spectrum excited by Reynold stresses is expected to peak around the convective turnover frequency while penetrative convection excites all low-frequency waves.

Generally, the excited low-frequency progressive waves are damped by the high thermal diffusivity present in the ambient medium. The dissipation happens at different layers depending on the wave frequency and degree $l$, and as they dissipate, the waves also release their energy and angular momentum, thus modifying the stellar interiors properties.
However, {if the damping of a wave is weak enough}, it can undergo reflection {leading} to the formation of a standing gravity mode which may then be observed via asteroseismology. The damping of the wave strongly depends on its wavelength \citep[\textit{e.g.}][]{Zahn1997}. We can thus define a cutoff frequency $\omega_c$ below which the waves are damped too quickly to be able to form standing gravity modes. \cite{Alvan2015} and \cite{Ahuir2021} give 
\begin{equation}
    \omega_c (r)= \left([l(l+1)]^{3/2}\int_{r}^{R_{\rm rad}} K_T \frac{NN_T^2}{r^3}dr'\right)^{1/4},
\end{equation}
where $K_T$ is the thermal diffusivity, and $R_{\rm rad}$ is the launching points of IGW, at the top of the radiative region in our case.

\begin{figure}
    \centering
    \includegraphics[width=0.45\textwidth]{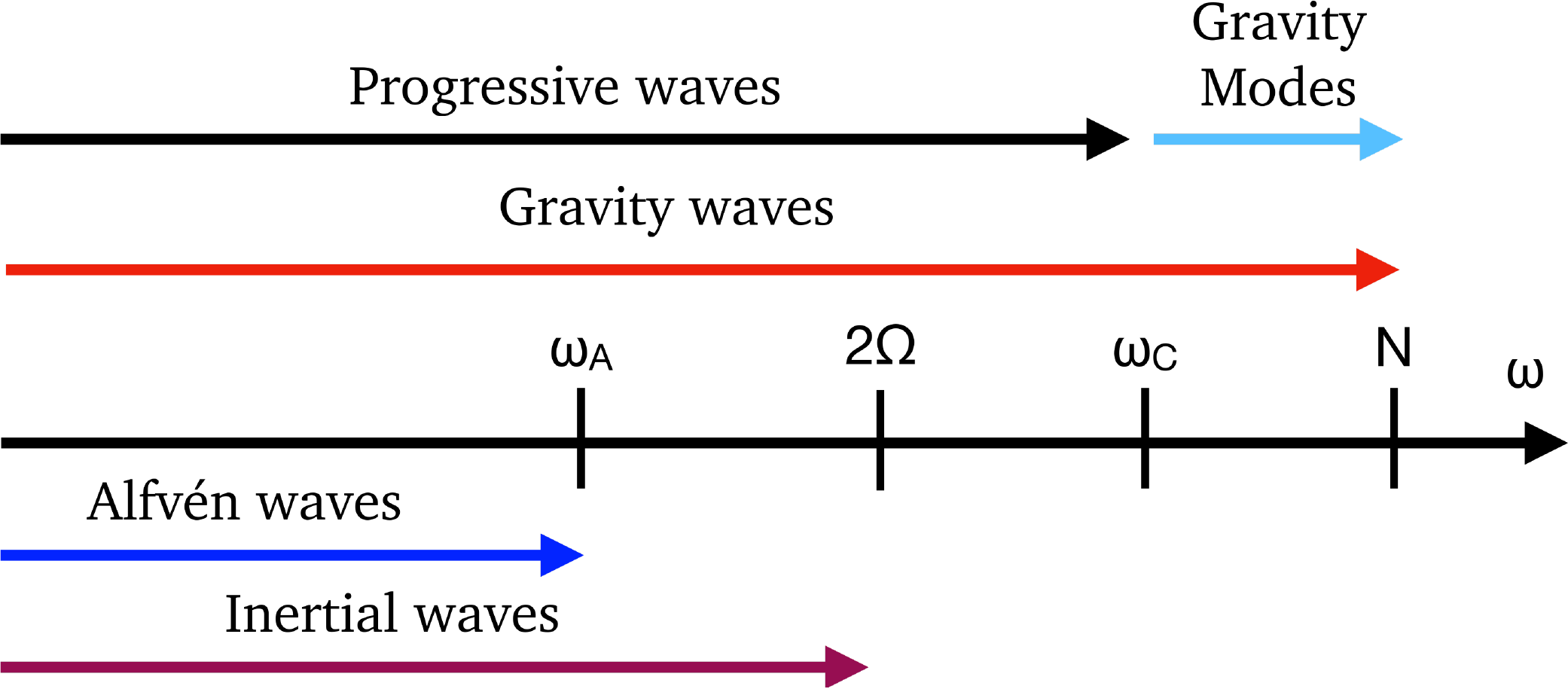}
    \caption{Wave spectrum in a magnetised rotating star}
    \label{fig:diagfreq}
\end{figure}

In a rotating star, we may also want to consider the formation of gravito-inertial modes \citep[][]{DintransRieutord2000,Prat2016,Prat2017,Prat2018} and progressive waves \citep{Mathis2009,Augustson2020} where the influence of rotation through the Coriolis acceleration is not negligible any more. This happens when a wave frequency is of the order of the characteristic rotation frequency ($2\Omega$) as indicated in figure~\ref{fig:diagfreq}. In this work, we focus on solar-type stars which are magnetically braked on their main sequence by the interaction of stellar winds with the large scale magnetic field \citep{Mattetal2015} ; they are therefore typically slow rotators and as a first step, {we {can} ignore the impact of rotation on gravity waves}.

\subsection{Tayler-Spruit Magnetic field}

The turbulent motions in {magnetised} radiative stellar interiors have been theorised to drive a small scale dynamo -- due to the {differential rotation}. It is born first from the growth of the {axisymmetric} azimuthal magnetic field driven by shearing flows associated to differential rotation \citep{Spruit1999}. {This $\Omega$-effect acts} on the large scale poloidal field building a toroidal field. The Tayler instability then onsets as the toroidal field becomes too large \citep{Tayler1973}. Finally, the electromotive force restores the poloidal field, closing the loop \citep{Zahn2007,Fuller2019}. This leads to the so-called Tayler-Spruit Dynamo \citep{Spruit2002}.\\
This dynamo can drive strong magnetic fields in the radiative region of stars and thus has important consequences for the transport of angular momentum in all type of stars at all stages of the evolution \citep[See \emph{e.g.}][]{Fuller2019,Moyano2022,Moyano2023}.   
{Recently}, \cite{Petitdemange2023} presented the first global numerical simulations displaying a dynamo with the theoretical TSD characteristics, in agreement with \cite{Spruit2002}'s formalism. 
Note that \cite{Fuller2019} also proposed a formulation from a slightly different dynamo loop which appears to be too efficient to agree with the results from the simulations by \cite{Petitdemange2023,Petitdemange2024} but comes in agreement with some of the simulations by \cite{Barrere2024}.

Here, we use a generalised version of the TSD to account for both the thermal and chemical stratification simultaneously as given by \cite{MaederMeynet2004}. We refer  the reader to {Appendix} \ref{Sect:App2} for the full description of the implementation of the TSD ; we will only recall {here} the expression of the Alfv\'en frequency in the radial and azimuthal directions.

In the TSD theory, the general Alfv\'en frequency is given by $\omega_A =$ B/$\sqrt{4\pi\rho}r$ with $\rho$ the local density. The magnetic field generated by the TSD has a radial ($B_r$) and an azimuthal ($B_\varphi$) component with different strengths. In particular, one can show from \cite{MaederMeynet2004} that the Alfv\'en frequencies associated with {local azimuthal and radial }fields are 
\begin{align}
\label{Eq:omal_comp}
    \omega_{A,\varphi} = \omega_{A}  \frac{m}{\sin\theta} && \textrm{ and } && 
    \omega_{A,r} = \frac{\omega_A^2}{N_{\rm eff}}\left(\frac{N^2}{\omega^2}-1\right)^{1/2} \sqrt{l(l+1)}
\end{align}
with the reduced Brunt-V\"ais\"al\"a frequency $N^2_{\rm eff} = \frac{\eta/K_T}{\eta/K_T+1}N_T^2 + N_\mu^2$. The full derivation is shown in Appendix B, {where we have assumed $N\gg\omega_A$, which is generally the case in stellar radiation zones}.

Finally, since $\omega_{A,r}$ depends on the frequency, we can define a critical value $\omega_{A,r,c}$ where $\omega=\omega_{A,r}$. With {again} the assumption of a strong stratification{ ($N\gg\omega_A$, $N\gg\omega$)} it comes \citep{Fuller2015} :
\begin{equation}
    \omega_{A,r,c} \simeq \omega_A \sqrt{\frac{N}{N_{\rm eff}}}\left[l(l+1)\right]^{1/4}.
\end{equation}

\subsection{Interactions between IGW and magnetic field}
\label{Sect:Interactions}
\subsubsection{Observable signature}
\cite{Garciaetal2014} found in the \textit{Kepler} data a population of red giant stars with reduced or even missing {mixed} modes in their pulsation spectrum that they associated to the presence of a strong field in the core. \cite{Fuller2015} showed that part of the low-frequency IGW can convert into Alfv\'en waves when the strength of the magnetic field is high enough. This should lead to observable consequences on the spectrum of strongly magnetised stars. For example, based on this  work, \cite{Stello2016} found that the presence of a magnetic field in  the core of red giants was a strong function of the stellar mass, concluding that the magnetic field was generally a remnant of a convective core dynamo during the main-sequence of intermediate-mass stars. 
More recently, \citet{Bugnet2021}  (see also \cite{Loi2021,Mathis2021,Bugnet2022,MathisBugnet2023,Rui2024,Lignieres2024}) computed theoretical asteroseismic signature of a stable axisymmetric and non-axisymmetric fossil field on the stellar oscillation frequency spectrum, allowing to {estimate} the magnetic field strength in red-giant interiors. The exercise was performed by \cite{Li2022,Li2023}, \cite{Deheuvels2023}, and \cite{Hatt2024} to detect magnetic field of several tens of thousands of Gauss in the core of red-giant stars and by \cite{Lecoanet2022} in the case of the main sequence massive star HD43317 with a field of the order of $5\times10^5$G.

For main sequence low-mass stars, {IGW} are generally progressive and produced at low frequencies \citep[e.g.][]{Alvan2014}. However, \cite{Lecoanet2017} and \cite{RuiFuller2023} showed that when the ambient magnetic field is large enough, gravity waves become evanescent and convert to magneto-gravity waves when their local frequency goes below that of the Alfv\'en frequency. The mechanism is complex and requires to reach high orders {in the WKB development or to consider modes with non-harmonic time dependences} to fully capture the subsequent spectrum \citep{RuiFuller2023}. 
 
\subsubsection{Angular momentum transport}
The Coriolis and Lorentz forces modify {the propagation cavity of the waves and trap them} in some parts of a given radiative region. 
In particular, the consequences on the transport of angular momentum by gravity waves modified by rotation and a toroidal field have been studied by \cite{MDB2012}. Following the approach of \citet{TalonCharbonnel2003}, they define a parameter $\mathcal{P}_m$ \citep[Eq. 118 in ][]{MDB2012} to evaluate the fraction of the total {momentum} flux that is transmitted {to waves at the convective-radiative boundary surface, for a fixed excitation rate. Note that in reality, the excited flux is also modulated by the rotation and magnetism \citep{Augustson2020,Bessila2025}. In practice, $\mathcal{P}_m$ gives the fraction of the spherical surface at the convective boundary where magneto-gravito-inertial waves are propagative and can be excited. The effective surface where the transmission occurs is reduced due to the combined effects of the equatorial trapping caused by the Coriolis force, and both vertical and equatorial trappings induced by the Lorentz force}. The $\mathcal{P}_m$ function writes:
\begin{equation}
\mathcal{P}_m = \left(1-\frac{m^2\omega_A^2}{\omega^2}\right)\frac{1}{\textrm{max}\left(1,\nu_{M,m}\right)},
\label{eq:Pm}
\end{equation}
with  
\begin{equation}
\nu_{M,m} = \frac{2\Omega}{\omega}\left(1-m\frac{\omega_A^2}{\Omega\omega}\right) \frac{1}{1-\left(\frac{m\omega_A}{\omega}\right)^2}
\end{equation}
as the control parameter of the system. Note also that as $\omega < |m|\omega_A$, the waves do not propagate any more. 
{Although this function was derived for a purely toroidal field, it was shown to give similar trends to the cases with more complex field topologies, which would be closer to realistic conditions \citep{Dhouib2022,Rui2024}.}\\
{For a fixed excitation rate, $\mathcal{P}_m$ thus qualitatively estimates the angular momentum flux ratio transmitted at the convective boundary, comparing the rotating case with a toroidal magnetic field with the non-rotating, non-magnetised case. 
\citet{MDB2012} also showed that the radiative damping is modified by the Coriolis acceleration and the Lorentz force, and thus the radii at which the internal waves are damped become closer to the excitation region. 
As a first step, the present work therefore focuses on how the latitudinal extent of the cavity of propagation of the waves at the convective/radiative interface affects the transmission of angular momentum from the convective envelope to the radiative core. The evaluation of the role of rotation and magnetism on the radiative damping and on the excitation rate is therefore beyond the scope of this paper.   
We assume that the transmitted angular momentum flux is deposited before reaching the centre of the star, which provides an estimate of the total torque applied on the radiative core at a fixed excitation rate.}
\\
{In the following section, we assess the impact of a TSD-generated magnetic field on the propagation of gravity waves in a solar-mass star, with particular focus on the angular momentum transport using this formalism. }

\section{Case of a 1 M$_\odot$ star through its evolution}
\subsection{The stellar evolution model}
\label{Sect:App1}
To compute the structure and the rotational evolution we use the 1D {stellar structure and evolution} code STAREVOL, which {self-consistently} computes the evolution of angular momentum \citep[See][for a more detailed description of the code]{Siess2000,Palacios2003,DecressinMathis2009,Amard2019}.
The composition follows \cite{AsplundGrevesse2009}, with \cite{KrishnaSwamy}'s fit for the atmosphere.
The mixing length $\alpha_{MLT}=2.11$ comes from a solar calibration of a non-rotating model.
The transport of angular momentum is described with the formalism of \cite{Zahn1992} and \cite{MathisZahn2004}, with the horizontal turbulence --along the isobars-- being much more important {than the vertical one} {because of the anisotropy of the turbulent transport triggered by the stable stratification \citep{Zahn1992,Mathis2018}}. We treat the meridional circulation as an advective process and all other mechanisms as diffusion terms according to the following equation
\begin{equation}
    \rho \frac{{\rm d}}{{\rm d}t}\left(r^2\Omega\right)= \frac{1}{5r^2}\ddr{} \left(\rho r^4 \Omega U_r\right) + \frac{1}{r} \ddr{} \left(r^4\rho \left(\nu_{v,v} + \nu_{TS}\right) \ddr{\Omega}\right),
\label{eq:amevol}
\end{equation}
where $\rho$, $ r$, $\nu_{v,v}$ and $U_r$ are the density, radius, vertical component of the turbulent viscosity, and radial component of the velocity of the $l=2$ meridional circulation on a given isobar, respectively. We implemented for the first time in STAREVOL the transport by the TSD as described in Appendix~\ref{Sect:App2} to compute the corresponding diffusivity $\nu_{TS}$.
{We recall the reader that both the rotation profile and the magnetic field properties are strongly determined by the chosen prescription, hereby \cite{Spruit2002}.}
{Finally, the external torque is computed following \citet{Mattetal2015}, calibrated to reach the solar surface rotation rate at the age of the Sun, and a saturation of the dynamo-generated magnetic field for $\frac{P_{\rm rot,\odot}/\tau_{c,\odot}}{P_{\rm rot}/\tau_c} < 0.1$, with $\tau_c$ and $P_{\rm rot}$ the convective turnover timescale at $H_P/2$ above the base of the convective region and the surface rotation period, respectively. Finally, we acknowledge that most observables tend to show a weakening of the surface magnetic torque beyond roughly the age of the Sun for solar-like stars \citep{Vansaders2016,David2022,Bhalotia2024,Metcalfe2025}. However, we do not expect this to qualitatively affects the results of this paper. Thus, to avoid adding any more parameters to the model, we kept the same expression for the stellar wind torque during the whole evolution.}

    \begin{figure}
        \centering
        \includegraphics[width=0.35\textwidth]{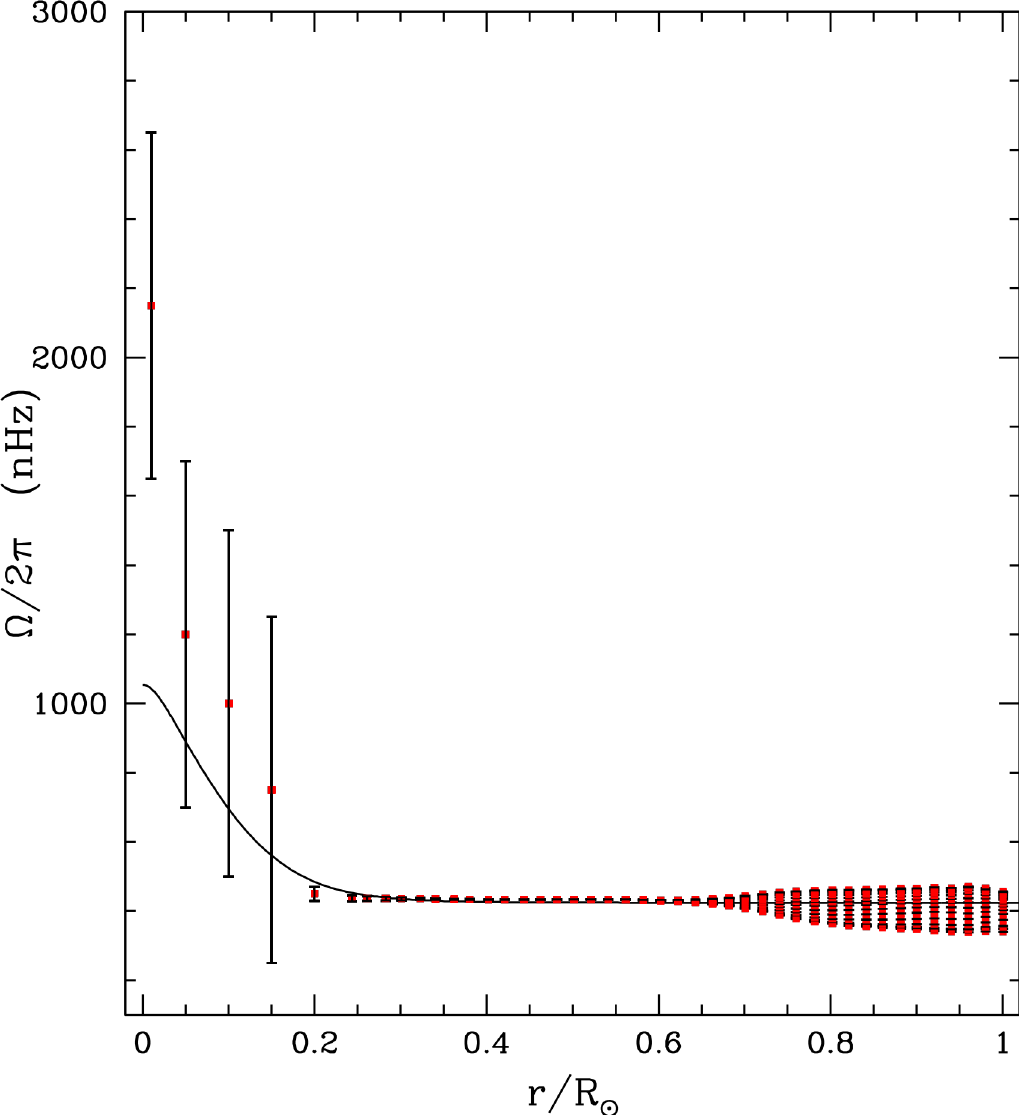}
        \includegraphics[width=0.4\textwidth]{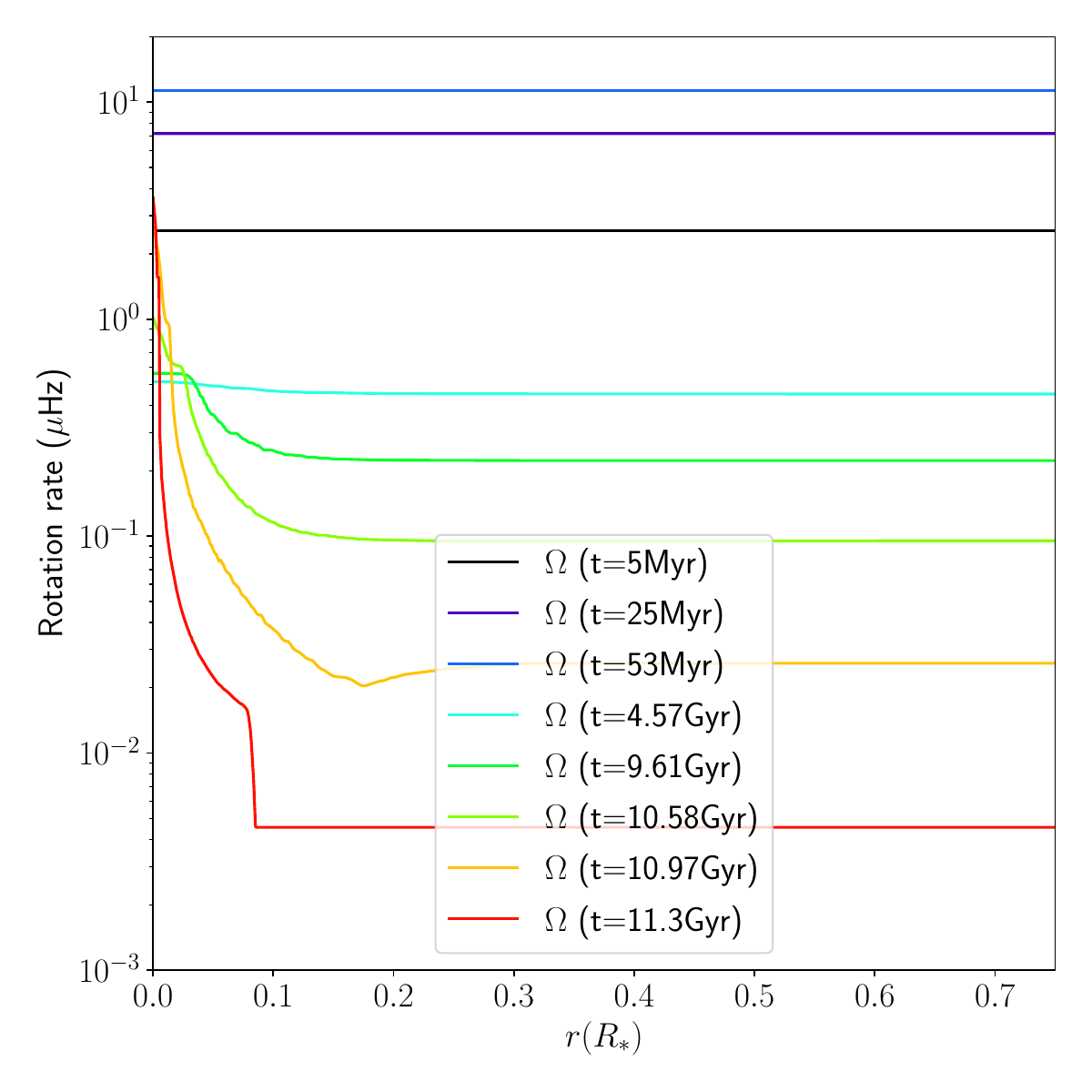}
        \caption{{Top. }Rotation profile at the age of the Sun for the 1M$_\odot$ model. The points and their error bars are from \cite{EffDarwich2008}. {Bot. }Rotation profile evolution of the 1M$_\odot$ model through the ages as indicated on each plot.}
        \label{fig:profrot_all}
    \end{figure}

We run a 1 M$_\odot$ evolution model with rotation from the early pre-main sequence to the red giant branch. In this section, we present {the} {key} moments of the stellar life to represent the evolution of the structure and various cavities crossed by typical {IGW} excited at the base of the convective envelope. {We also compute the fraction of angular momentum that is transmitted by internal waves from the convective envelope to the radiative core using equation~(\ref{eq:Pm}).}

In figure~\ref{fig:allfreq}, the red line indicates $\omega_{TO}$ the convective turnover frequency at the base of the convective envelope. The wave excitation is expected to be maximal around this value \citep{Press1981,Fuller2014}. In particular, during the pre-main sequence and after the TAMS, the convective eddies are much larger and the peak excitation happens at lower frequencies.
We can distinguish two classes of waves separated by the cut-off frequency ({blue shaded region} on \ref{fig:allfreq}). Above this value, the waves {form standing g-modes}. Below the cut-off frequency, {the waves are progressive and will dissipate due to thermal diffusion \citep{Alvan2013}, critical layers, or non-linear breaking \citep{Zahn1997,Mathis2025}.} 
The evolution of the rotation profile under the influence of both hydrodynamical processes {(meridional circulation and shear instabilities)} and the TSD is presented in figure~\ref{fig:profrot_all}. We first confirm the findings of \cite{Eggenberger2005,Eggenberger2019a} that during the pre-main sequence, despite the contraction, the star is very close to solid-body rotation due to the fast rotation and the low stratification of the star which both efficiently trigger the TS instability. It is only during the main sequence, a little before the age of the Sun that a differential rotation sets in close to the core where the stratification has increased enough to prevent the TS instability. Finally, as the 1M$_\odot$ star evolves to the sub-giant branch (SGB) and rises along the red giant branch (RGB), the differential rotation increases as neither the TS dynamo nor the hydrodynamical instabilities are efficient enough to extract the concentration of angular momentum in the core associated to the contraction \citep{Cantiello2014}. 

In figure~\ref{fig:allfreq}, we display the frequencies relevant to the characterisation of the wave properties as described in Sect.~\ref{Sect:Physics} along the evolution of the 1M$_\odot$ star. The top three panels present the pre-main sequence structure evolution, the central left one corresponds to the Solar age, the following ones show the evolution from the terminal-age of the main sequence to the red giant branch stage as presented in the top left panel of figure~\ref{fig:allfreq}.
Overall, we first verify that we are always in a highly stratified situation, which means that we are always dominated by the Brunt-V\"ais\"al\"a frequency (black line in fig~\ref{fig:allfreq}). 
The cut-off frequency $\omega_c$ {delimits the regime of stationary {g-modes }modes ({blue shaded region} in fig~\ref{fig:allfreq}) from the one of progressive {IGW}}. Its value remains below the Brunt-V\"ais\"al\"a frequency, therefore both are able to propagate in the radiative region through the evolution.

\subsection{Standing gravity modes}
Standing gravity modes are built by the constructive interference of {IGW}. They are found at frequencies higher than the cutoff frequency $\omega_c$. Since $\omega_c \geq \omega_{A}$ for the entire evolution depicted here, the modes are not affected by the magnetism generated by the TSD. This goes along with the conclusions of recent work by \cite{Fuller2015} and by \cite{RuiFuller2023} that only strong magnetic fields will affect the observable modes. This enforces the conclusion that the magnetic field capable of hindering/retaining/trapping some of the modes in the core of red giant stars comes from a fossil magnetic field and not a TSD mechanism \citep{Bugnet2021,Li2022}. 

\subsection{Progressive waves}
Progressive waves can follow several possible paths depending on their excited frequency.
During the pre-main sequence, the fast rotation ensures a strong TS dynamo and thus a high Alfv\'en frequency $\omega_{A}$ of a few tenths of microHz from the base of the convective envelope to the centre of the star (or the edge of the convective core when present). All the gravity waves excited with frequencies below this value are quickly converted into magnetogravity waves (MGW) and keep this identity as they propagate to the core. They may locally enhance the torque associated with the magnetic field through Maxwell stresses, here convened by the TSD. 
All higher frequencies propagate as IGW and deposit their angular momentum when {the thermal diffusion dissipates them, or where they break or encounter a critical layer }\citep[See \textit{e.g.}][]{TalonCharbonnel2005,Alvan2013,Charbonnel2013,Mathis2025}.
{At the ZAMS, the 1M$_\odot$ star has a convective core and the strong chemical gradient above its surface prevents the TSD from maintaining a strong coupling. This locally allows for a fast rotation and a larger Alfv\'en frequency.}

On the main sequence, {the convective core has disappeared but the stratification starts building}, limiting the strength of the TSD and slowly increasing the differential rotation in this region while decreasing the Alfv\'en frequency below 10 nHz in the core.
As we get closer to the convective boundary, the stratification is less important, and the upper half of the radiative region is still driven by the torque from the magnetised stellar {wind. Therefore, these layers undergo a} higher shear that keeps the local magnetic field generated by the TSD at a high level. 
It translates into a higher radial Alfv\'en frequency and {thus }a larger part of the IGW spectrum excited at the base of the convective envelope can be directly converted in MGW.
{The interesting part is that, after crossing the strongly magnetised region where the TSD is very efficient, the waves may change again their identity as their frequency becomes higher than $\omega_A$. The exact becoming of these waves is not clear as of today and we keep the discussion for other specific papers. The waves of higher frequencies are still propagating as typical IGW.}

\begin{figure*}
   \centering
   \includegraphics[width=0.28\textwidth]{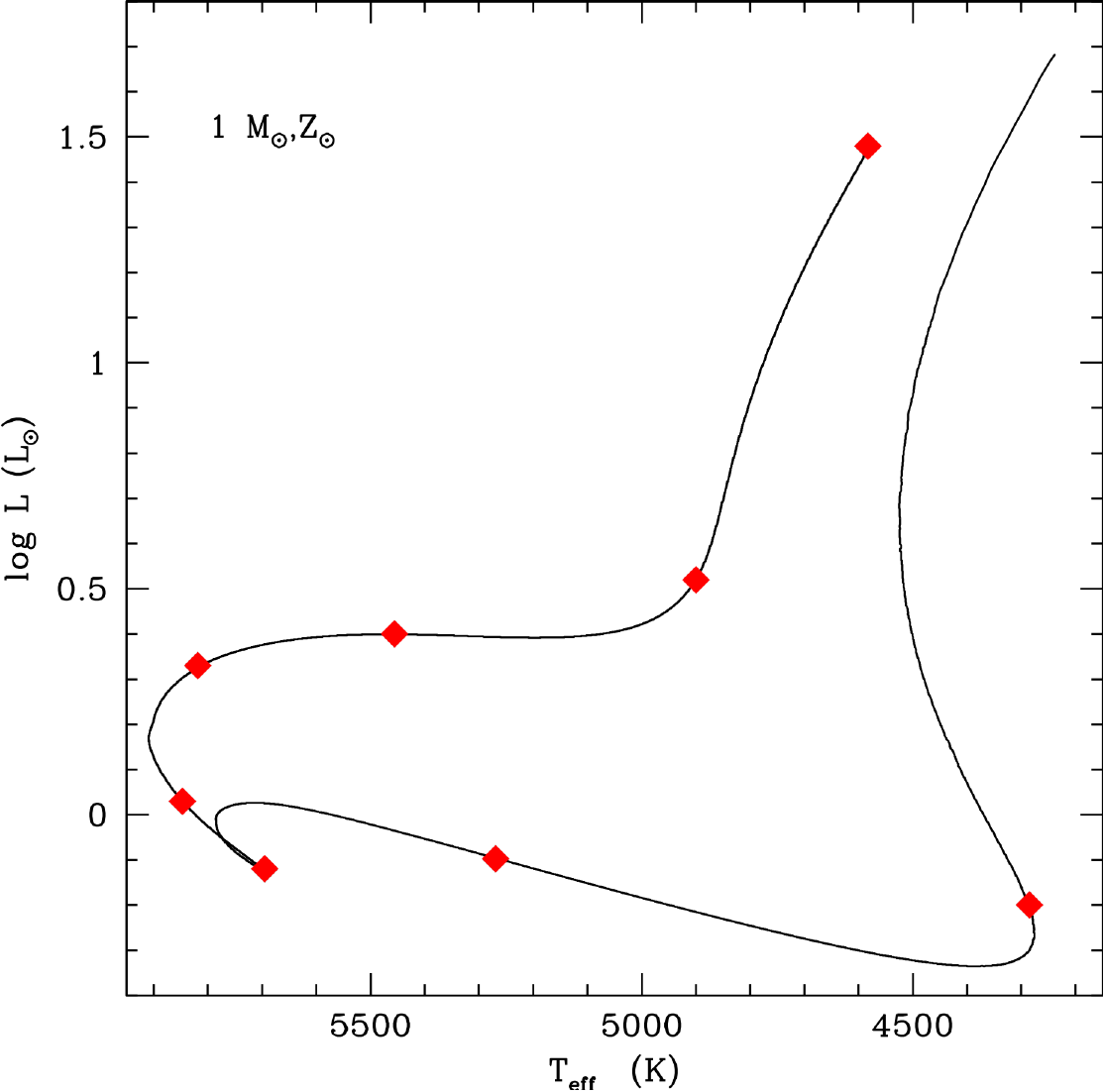}
   \includegraphics[width=0.32\textwidth]{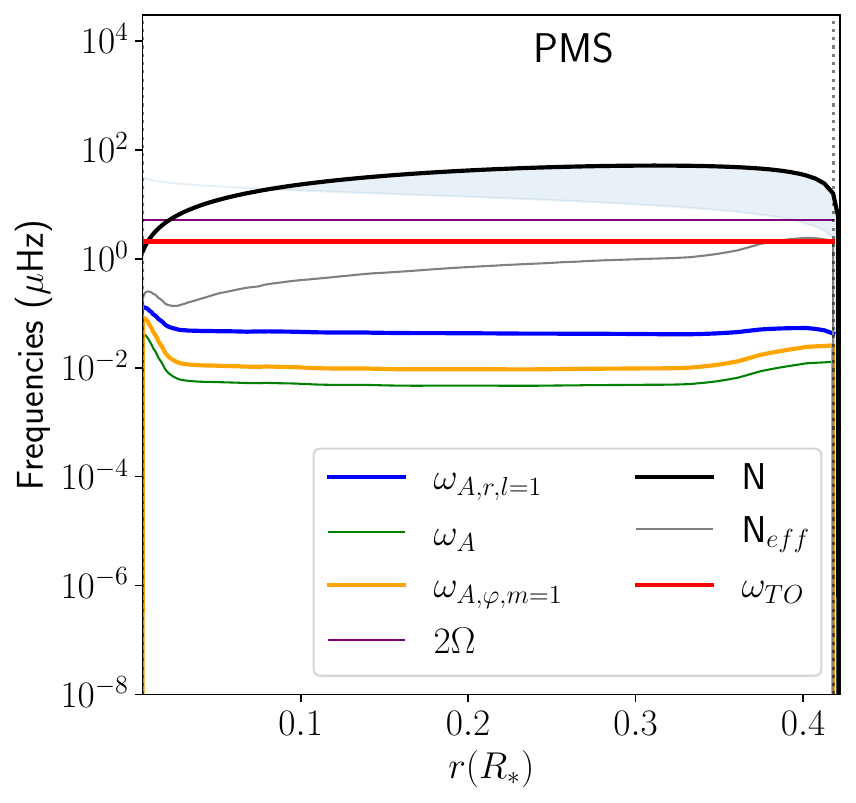}
   \includegraphics[width=0.32\textwidth]{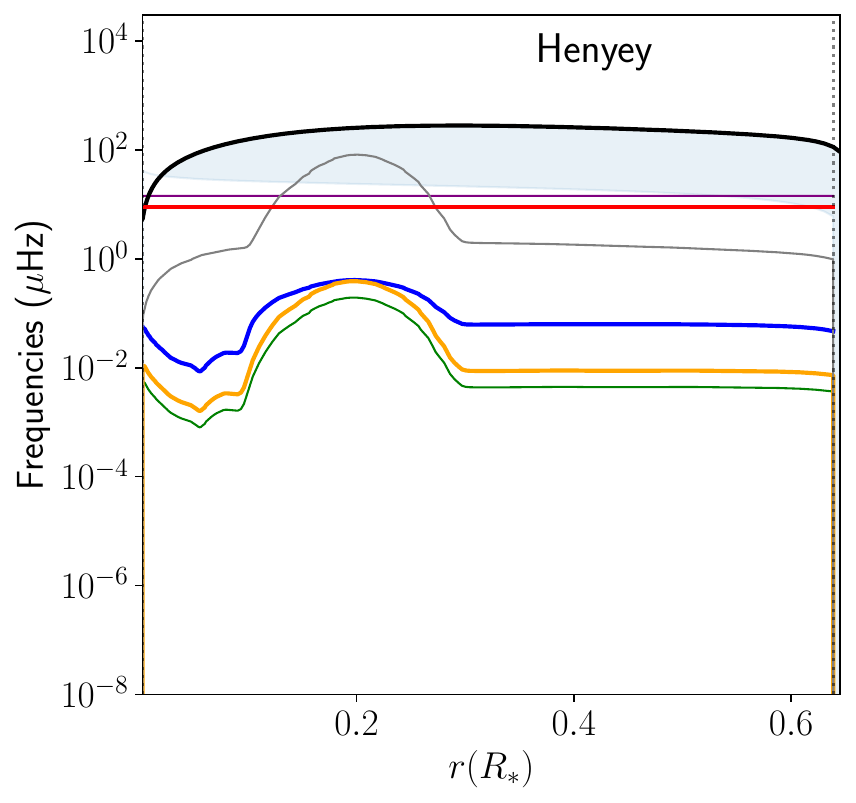}
   \includegraphics[width=0.32\textwidth]{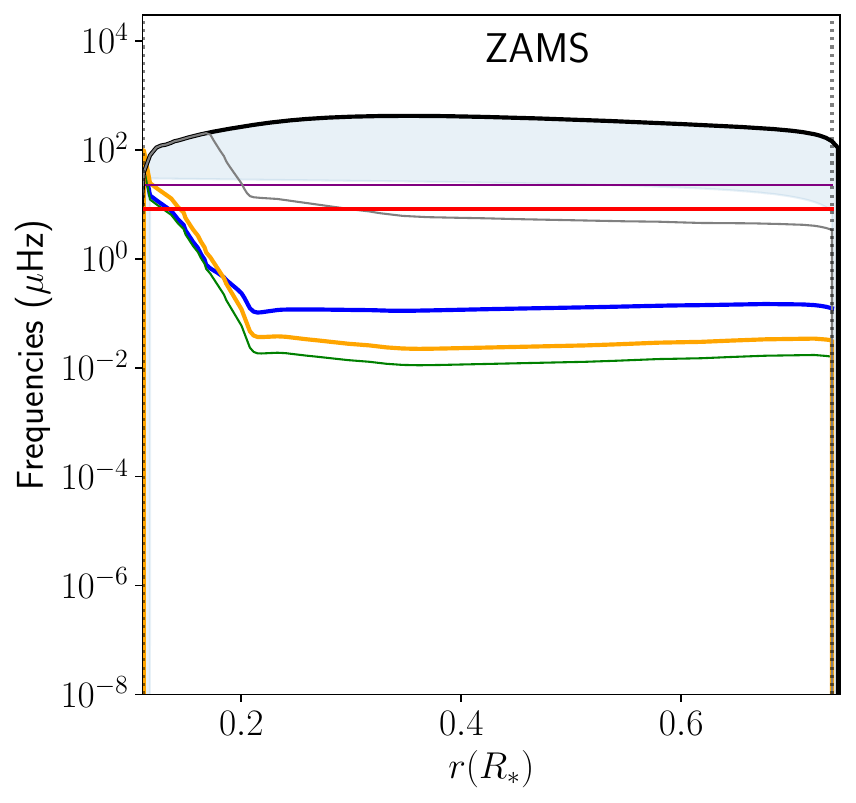}
   \includegraphics[width=0.32\textwidth]{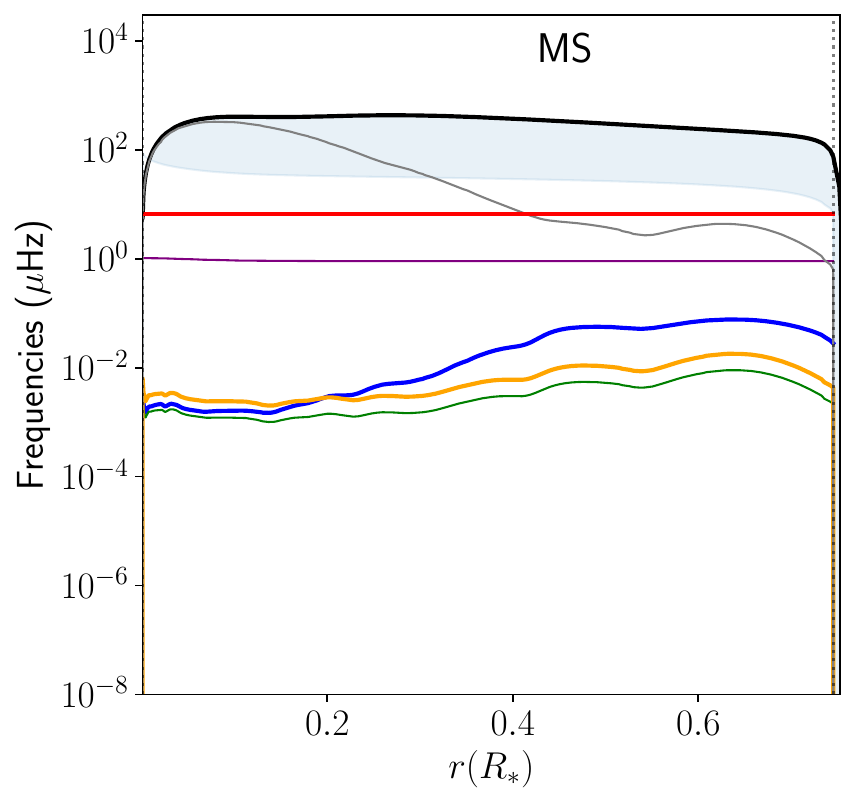}
   \includegraphics[width=0.32\textwidth]{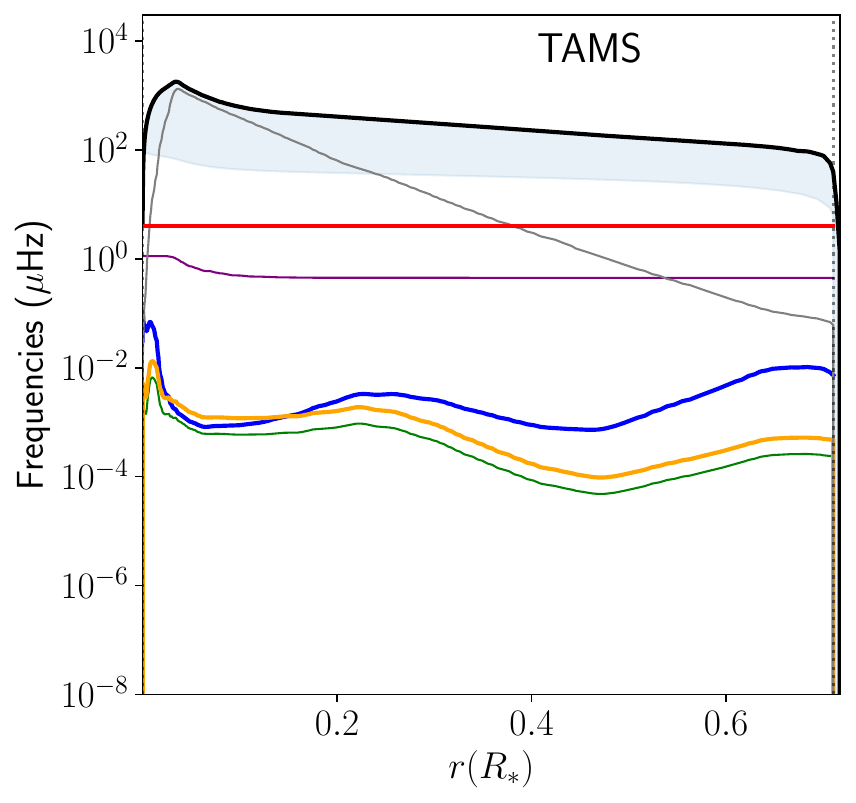}
   \includegraphics[width=0.32\textwidth]{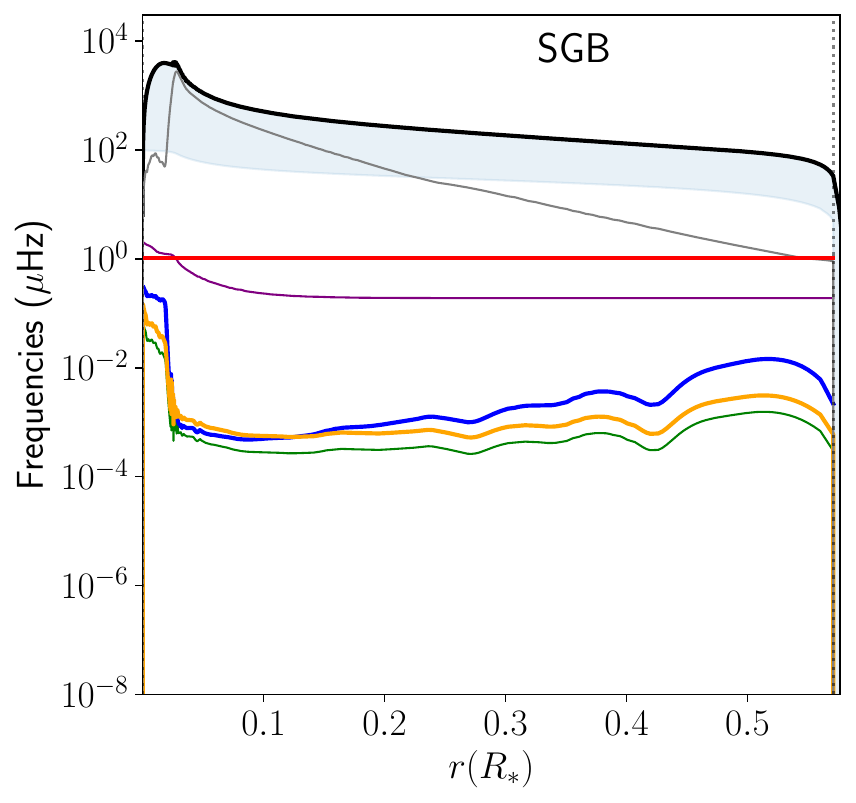}
   \includegraphics[width=0.32\textwidth]{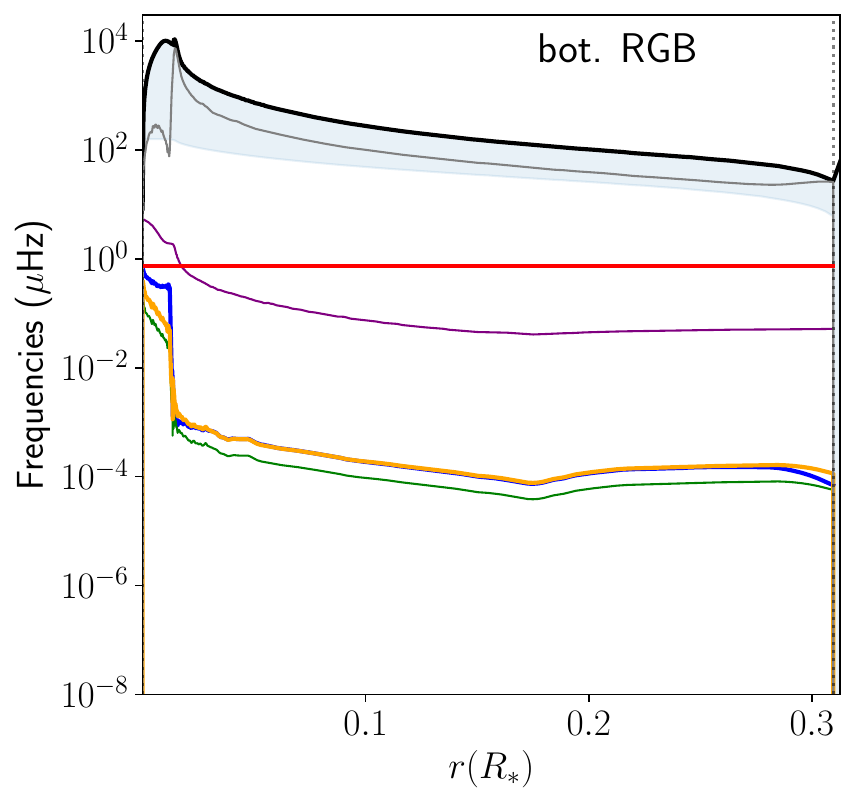}
   \includegraphics[width=0.32\textwidth]{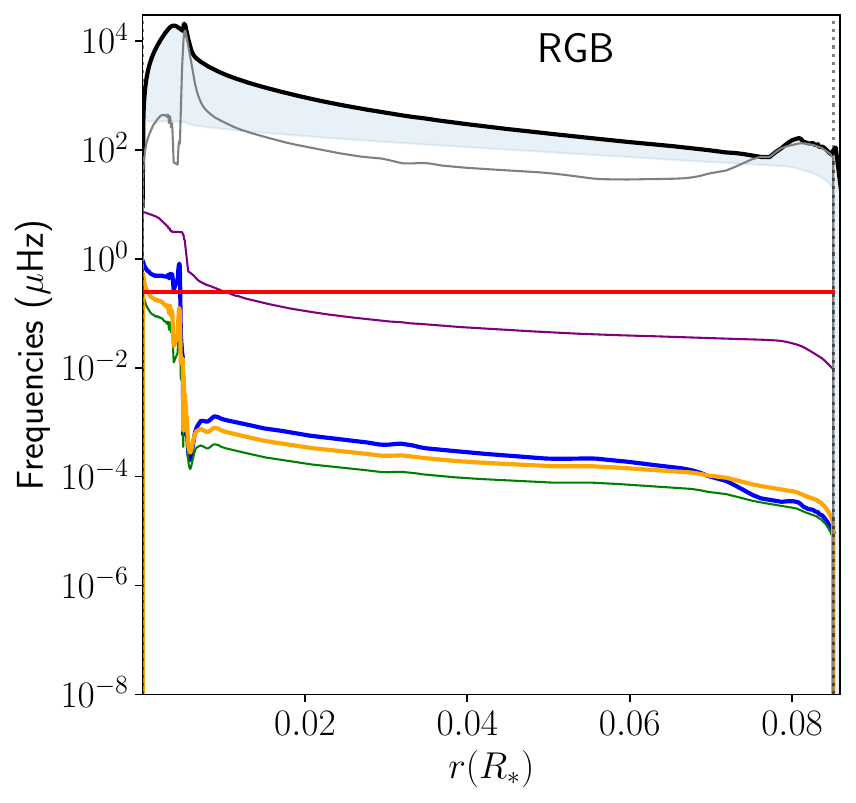}
   \caption{{Top left :} Location of each selected profile on the Hertzsprung-Russel diagram of the 1M$_\odot$ model. {The next diagrams show the v}ariations of the relevant characteristic frequencies along the evolution. From top center to bottom right : Early PMS (5 Myr), Henyey track (25 Myr), ZAMS (52 Myr), Solar age (4.57 Gyr), TAMS (9.61 Gyr), sub-giant branch (10.53 Gyr), base of the RGB (10.97 Gyr), and middle of the RGB (11.38 Gyr).  The light blue region indicates the frequency range of {standing }gravity modes. }
   \label{fig:allfreq}%
   \end{figure*}

As the star reaches the end of the main sequence, the stratification in the central layers has strongly increased and a chemical gradient has formed from the nuclear reactions. The combination of both stabilising factors locally hinders the TSD and reduces the transport of angular momentum, thus leading to a steep rotation gradient close to the core of sub- and red giants, in agreement with \cite{Vrard2022}. 
However, although the TSD is not sufficient to fully couple the central part to the rest of the radiative zone, the strong differential rotation and the faster rotation of the core are able to generate a strong magnetic field and more IGW frequencies can be converted to MGW. For example in the case of a star at the base of the RGB (central bottom panel on Fig~\ref{fig:allfreq}), IGWs with a frequency between 0.01-1 $\mu$Hz may be able to propagate all the way to the central region around $r=0.01R_*$ where their frequency becomes comparable to the Alfv\'en frequency.
{As discussed in the following subsection, the progressive waves that reach this region should locally release the angular momentum they acquired from the excitation region. They should therefore be accounted for in the transport of angular momentum as in \cite{Charbonnel2013} for example.}
The radial component of the magnetic field is generally associated with a conversion of IGW to MGW \citep{RuiFuller2023}, while the horizontal component is associated with a {vertical} magnetic trapping of the waves \citep{MDB2011}. In each diagram we show the radial and azimuthal components of the Alfv\'en frequency in blue and orange, respectively, as computed in the Appendix~\ref{Sect:App2}. {For most of the evolution}, the radial component dominates and low-frequency IGWs should preferably convert to MGWs. {At the ZAMS and during the main sequence}, as the core contracts, the azimuthal component becomes slightly higher than the radial one, and IGWs are likely to be {vertically} trapped as they reach the central regions. However, when the star goes towards the red giant branch, the effective BV frequency drops as the magnetic diffusivity increases, and $\omega_{A,r}$ rises again. 

On each diagram, the red lines indicate the turnover frequency, which generally corresponds to the peak of excitation by Reynold stresses for {IGW} \citep[See \textit{e.g.}][]{Fuller2014}. {We advise the reader to keep in mind that a whole spectrum of frequency is excited which also goes to lower frequencies. In addition, the spectrum excited by penetrative convection could be very important at lower frequencies \citep{Alvan2014,Pincon2016}.} 
It is clear from figure~\ref{fig:allfreq} that {along the main sequence} evolution, {the IGWs excited by the bulk of Reynold stresses have a characteristic} frequency much higher than the Alfven frequency and should therefore not be affected by the TSD magnetic field. {Again, low-frequency IGWs excited by penetrative convection could be affected by the TSD.} On the sub-giant and red giant branch, the convective envelope quickly deepens so that the convective turnover frequency decreases. Meanwhile, the strong increase of the stratification and the differential rotation in the core leads to higher magnetic field and thus to a higher $\omega_A$. The bulk of excitation is now comparable to the Alfv\'en frequency and most of the spectrum of excited IGWs can be converted to MGWs when the star climbs up the red giant branch. On the last panel for example, all the waves with a frequency below 1 $\mu$Hz are converted to MGWs.  
Note however that, in our simulations, the core rotation on the red giant branch is still too high compared to observations by almost an order of magnitude \citep{Cantiello2014}. Therefore, the shear is likely too important and the magnetic field generated as a consequence might be lower.

\subsection{Angular momentum transport}
\begin{figure*}
    \centering
    \includegraphics[width=0.31\linewidth]{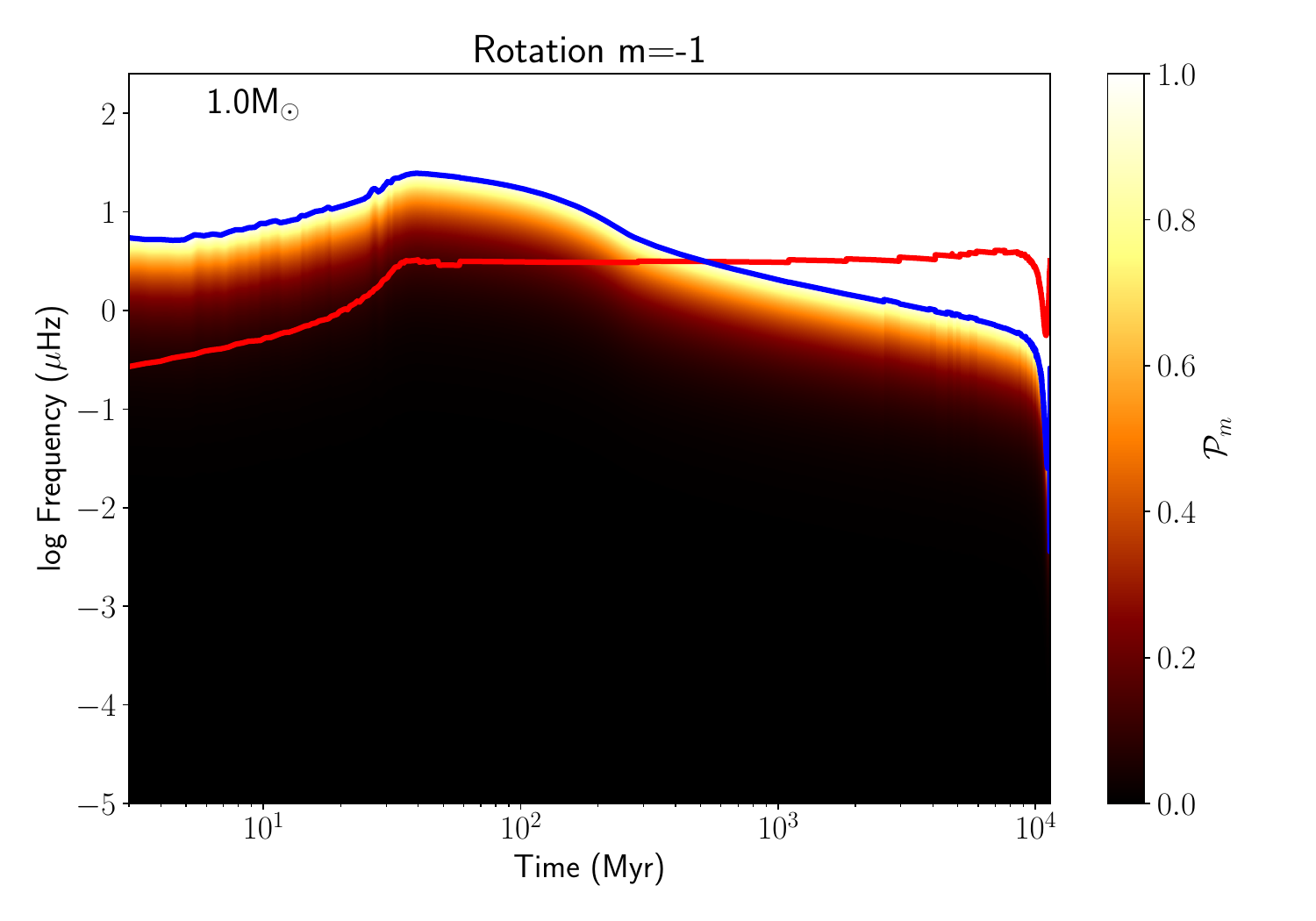}
    \includegraphics[width=0.31\linewidth]{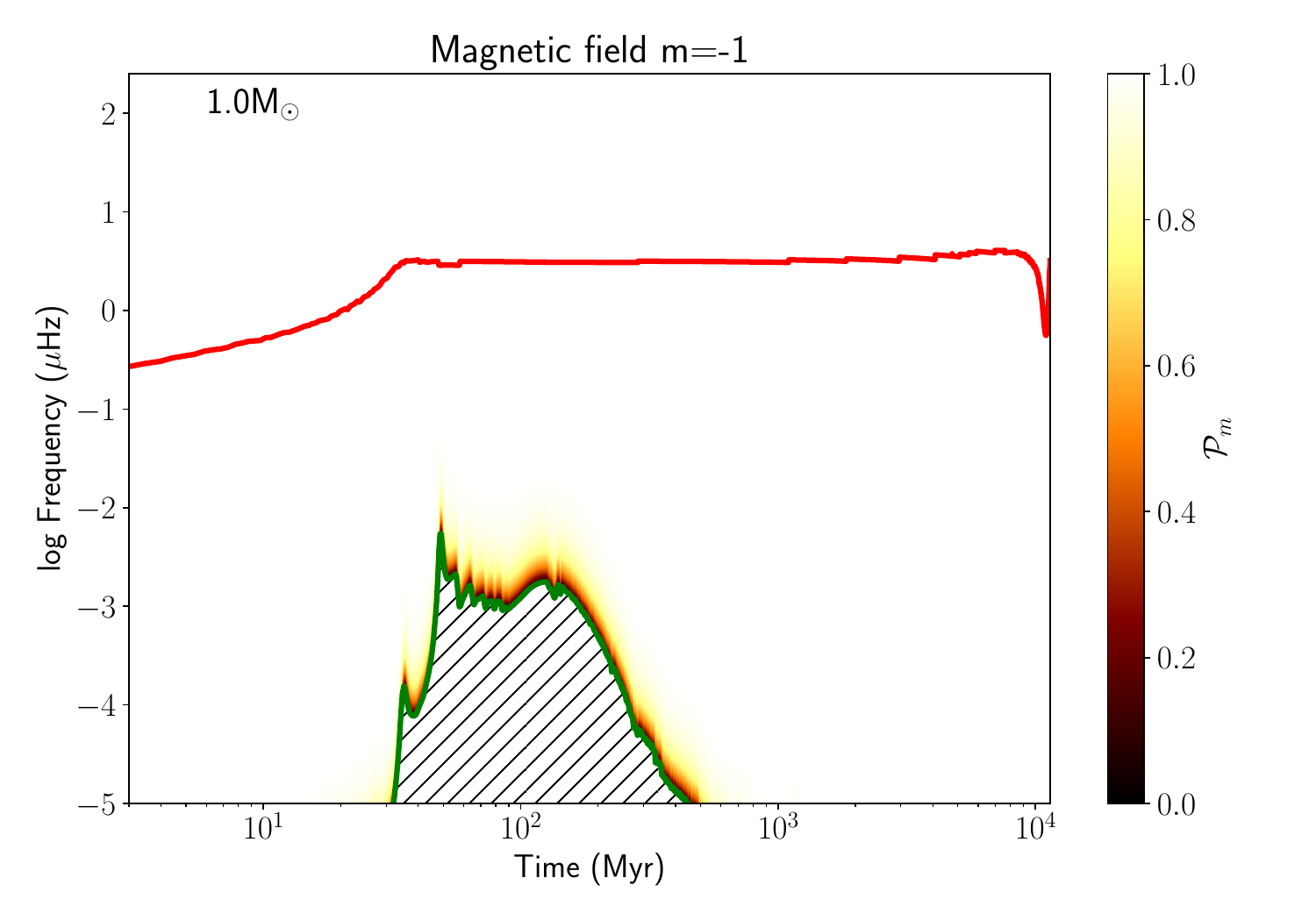}
    \includegraphics[width=0.31\linewidth]{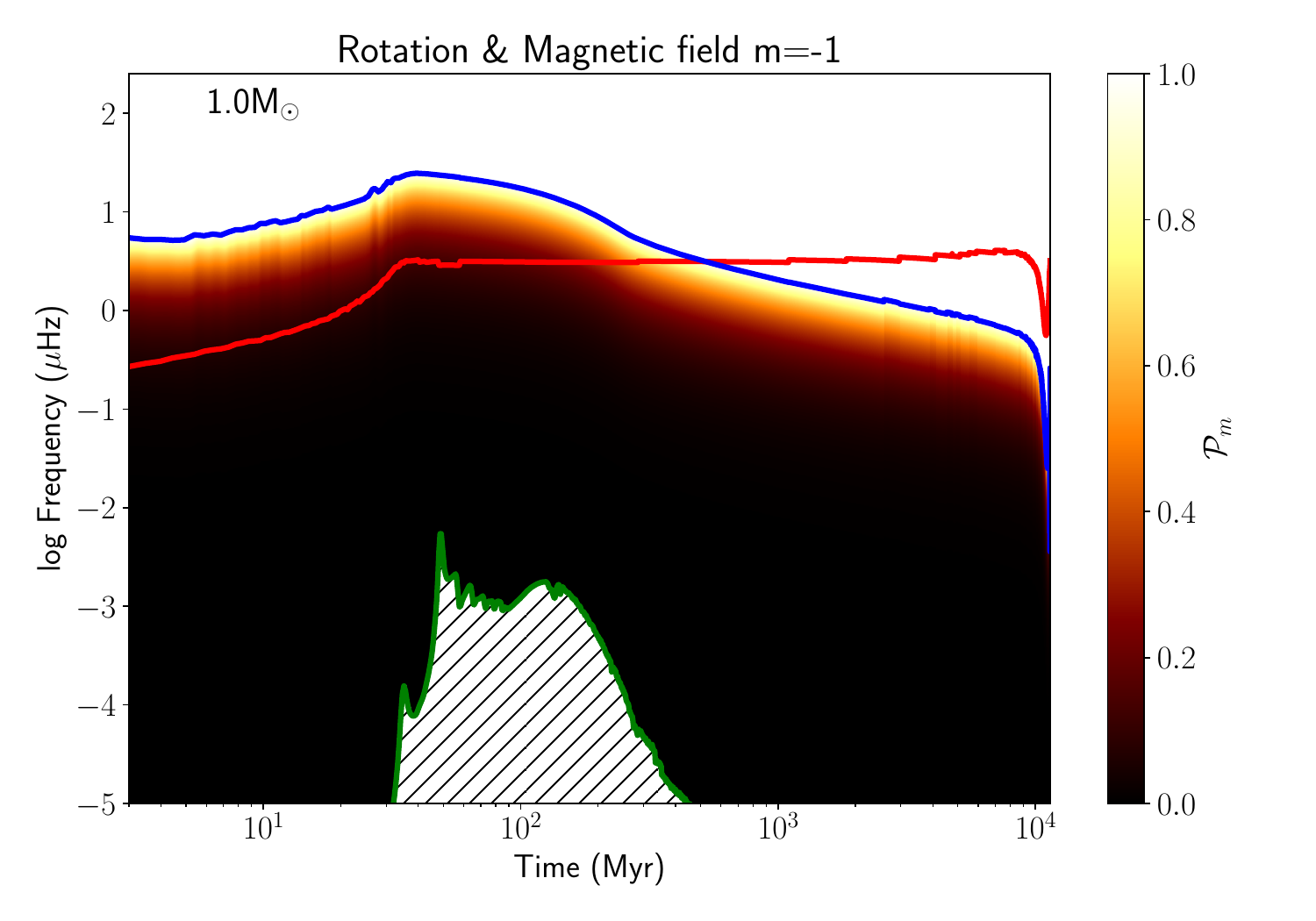}
    \caption{{Colour maps of the transmission function $\mathcal{P}_m$ from \cite{MDB2012} with $m=-1$ as a function of frequency and time for a 1M$_\odot$ star. The left (central) panel shows $\mathcal{P}_m$ when only the rotation (toroidal magnetism) is taken into account, the right panel presents the case with the full contribution of $\mathcal{P}_m$. 
    White regions show where the angular momentum flux carried by the wave is unaffected by the rotation and magnetic field while black regions indicate where it is. The dashed part of the diagram present the region where the waves are completely trapped vertically and therefore do not propagate. The red line shows the convective turnover frequency as a function of time, which acts as a proxy for the peak of excitation. The blue and the green lines are the inertial frequency ($2\Omega$) and the Alfv\'en frequency integrated over the upper 5\% of the radiative core, respectively.}}
    \label{fig:Pm_1Msun}
\end{figure*}

In a non-magnetised and non-rotating star, the wave excitation and the transport of angular momentum occur over the whole sphere. However, the rotation contains waves closer to the equator, and the magnetic field prevents the {full} transmission of angular momentum, both in the vertical and latitudinal directions. We take advantage of the work by \cite{MDB2012} to compute {the torque transmitted from the convective envelope to the radiative core in the rotating magnetised case when compared to the one in the non-rotating, non-magnetised one, for a fixed excitation rate}\footnote{{The effect of rotation and magnetic field on the excitation is not taken into account \citep{Augustson2020,Bessila2025}.}}. 
{The exact radius at which the waves are damped is determined mainly by their radiative damping, which is also modified by the Coriolis acceleration and the Lorentz force \citep{MDB2012}. Its evaluation for a given frequency power spectrum will be addressed in a forthcoming work.
In the current study we consider the magnetic field generated by TSD and the rotation profile in the upper 5\% in radius of the radiative core to act somewhat as a filter. We therefore compute the integrated $\mathcal{P}_m$ value over this thin layer to estimate the fraction of the flux of angular momentum that is transmitted to the radiative core.}
{Figure~\ref{fig:Pm_1Msun} shows the $\mathcal{P}_m$ factor defined in Sect.~\ref{Sect:Interactions} for $m=-1$ (large scale retrograde waves) as a function of the wave frequency and time for the 1M$_\odot$ star model under the influence of rotation (left), magnetism (center), and both rotation and magnetism (right).}

{The first notable element is the role of the rotation in the trapping, compared to the magnetic field. The left side figure, with rotation only, is indeed very similar to the right side figure where both the magnetic field and the rotation are taken into account. 
In practice, the TSD generated magnetic field only plays a role in trapping sub-alfvenic waves, which means for the current case, waves with a frequency below 0.01$\mu$Hz at most.
We therefore mainly focus on the role of rotation. We can see that, as soon as the waves become sub-inertial, the trapping is increasing. 
Note also that we only showed the case of retrograde waves where $m=-1$. The reason is that, although $\mathcal{P}_m$ depends on the sign of $m$ (Eq.~\ref{eq:Pm}), the difference between two $m$ of opposite signs varies with $\omega_A^2$. We therefore do not see any visible difference in $\mathcal{P}_m$ for prograde and retrograde waves.}

{We then compare the trapping region with the bulk of the excited spectrum of IGW, indicated by the red line in figure~\ref{fig:Pm_1Msun}.  During the early evolution, the combination of the larger envelope first with a low convective turnover frequency, and then the fast rotation leads to a very efficient trapping of the waves with frequencies lower than 7 $\mu$Hz during the early PMS, and 20 $\mu$Hz at ZAMS. Around 500 Myr, the star has spun down enough from the magnetised winds that most of the angular momentum carried by IGW can filter inward to the radiative core. We can therefore expect that angular momentum transport will be more efficient from this point on.}
{During the later evolution, on the one hand, the convective turnover frequency decreases as the convective envelope inflates in the subgiant branch and then the RGB, on the other hand, the rotation rate (and the Alfv\'en frequency) always remains much smaller, as can also be seen on the last two panels of figure~\ref{fig:allfreq}. This leads to an efficient transmission of the convectively excited angular momentum flux to the inner regions of the star. There, as previously discussed, the IGW can be converted to magneto-gravity waves close to the core, and thus the transport of angular momentum would be modified.}

\section{Other masses with radiative core}
{We present in this section the same evaluation on stars of masses between 0.6 and 1.2 M$_\odot$. Due to their faintness and their low oscillation amplitude, associated with the larger surface variability associated to magnetic activity,} the observable counterpart is much more challenging to obtain for stars with a mass lower than the Sun {\citep{Verner2011}. However, recent works with \emph{ESPRESSO} and TESS showed promising results for mid-K dwarves \citep{Campante2024,Hon2024} ; we therefore extended our study to this mass range.}

\begin{figure*}
    \centering
    \includegraphics[width=0.24\textwidth]{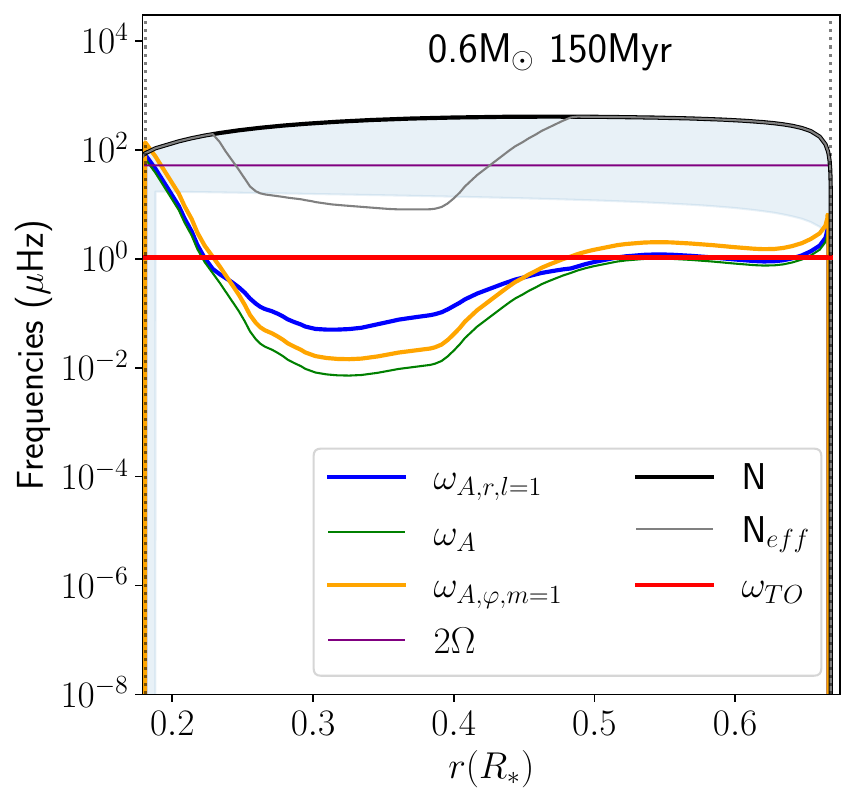}
    \includegraphics[width=0.24\textwidth]{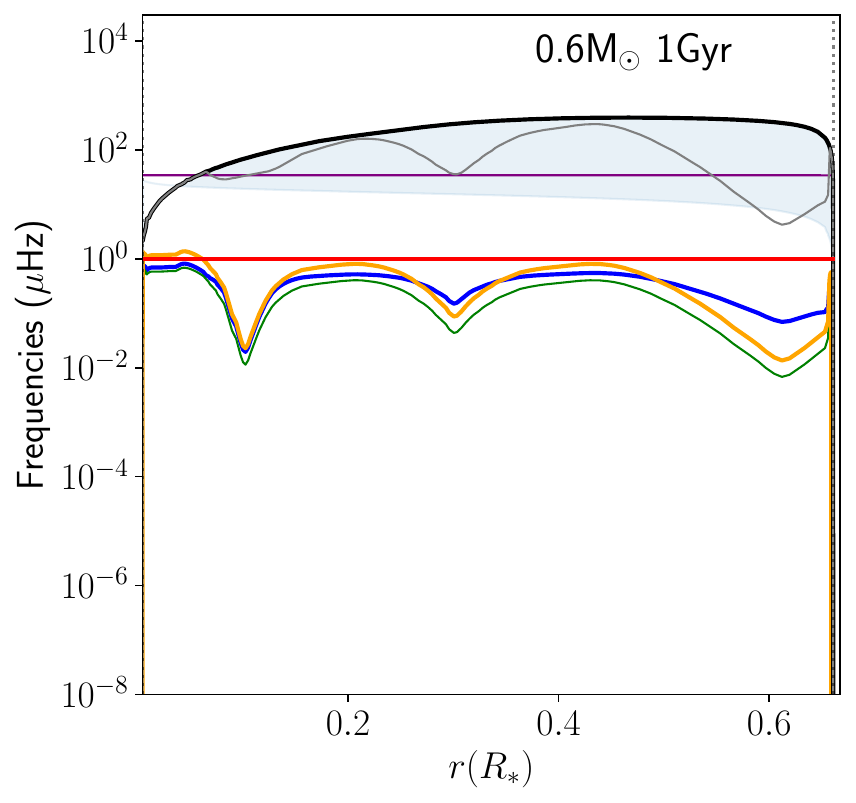}
    \includegraphics[width=0.24\textwidth]{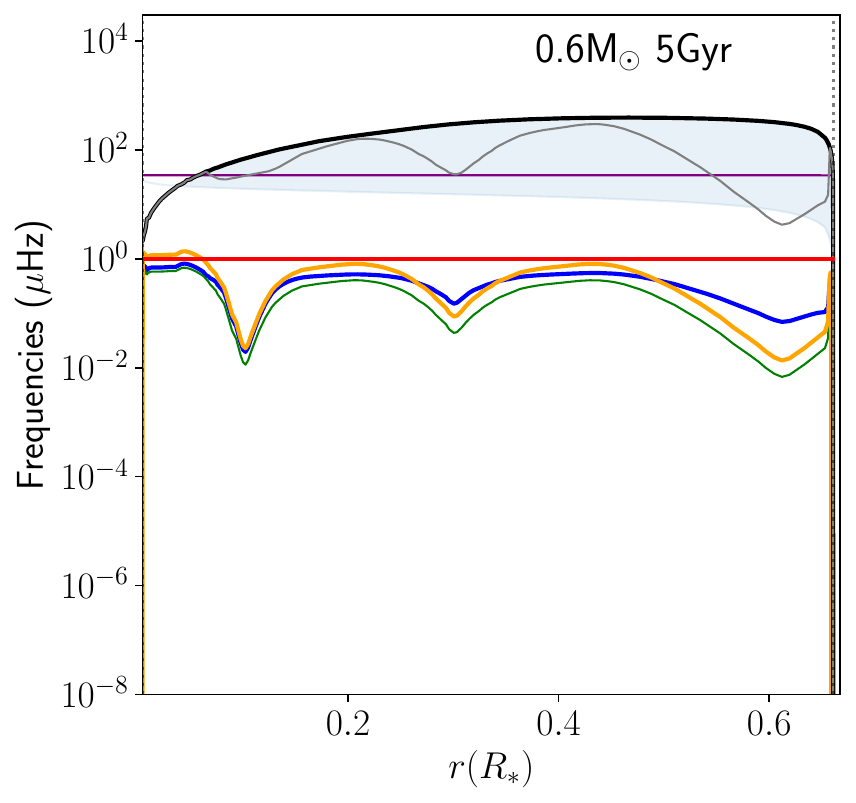}\\
    \includegraphics[width=0.24\textwidth]{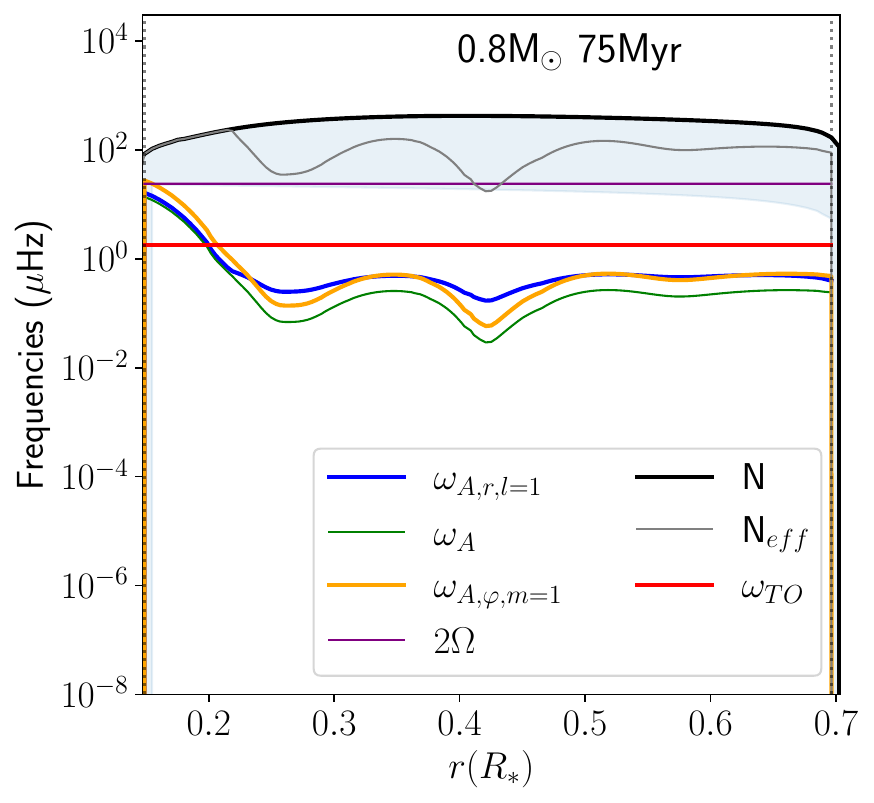}
    \includegraphics[width=0.24\textwidth]{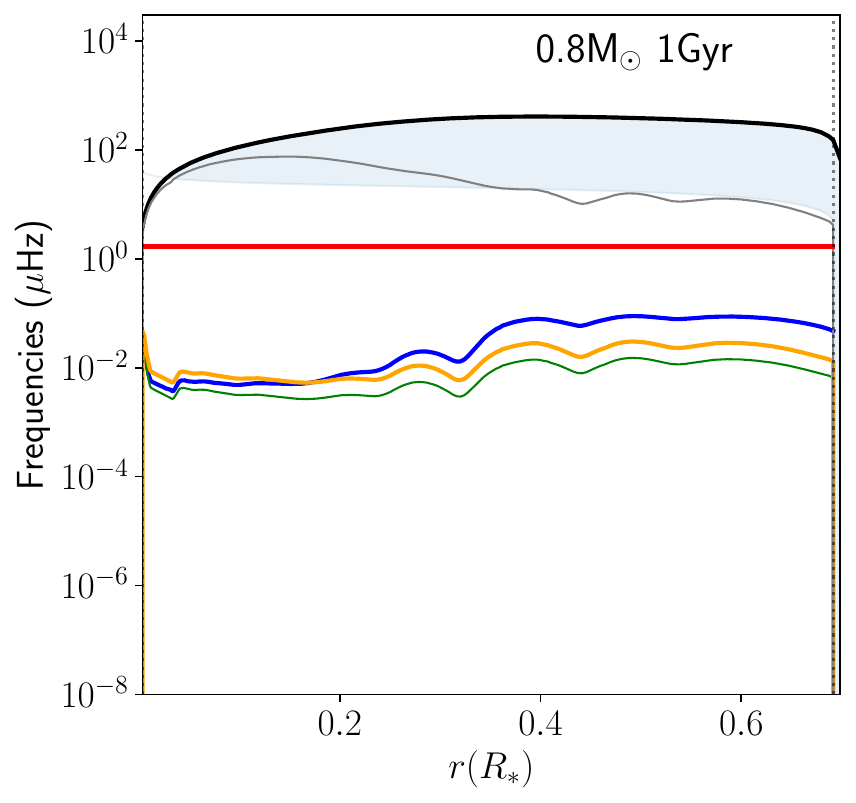}
    \includegraphics[width=0.24\textwidth]{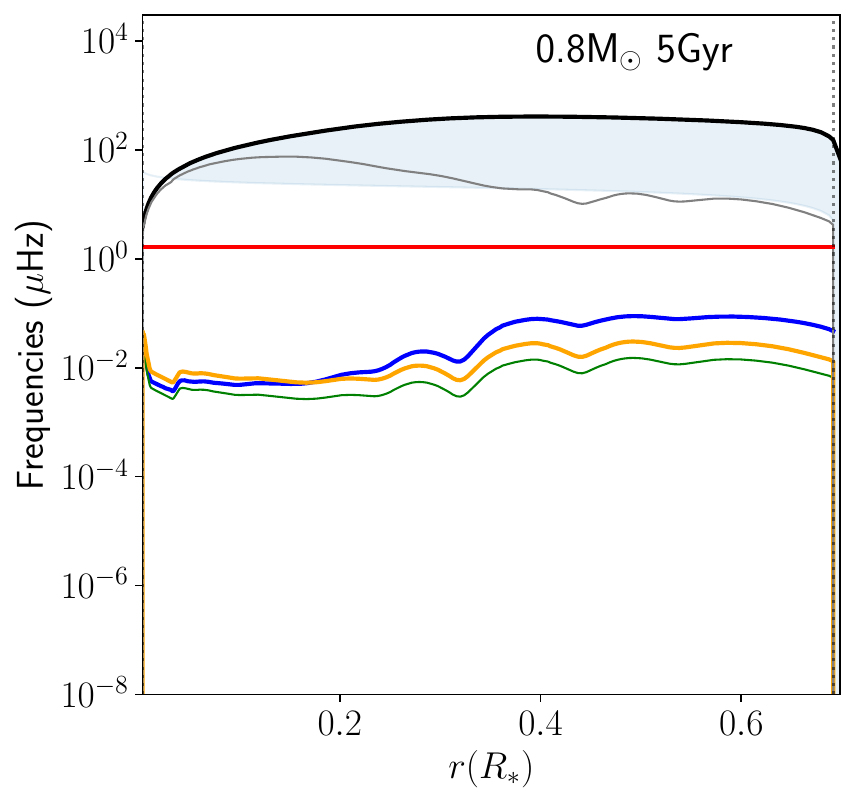}\\
    \includegraphics[width=0.24\textwidth]{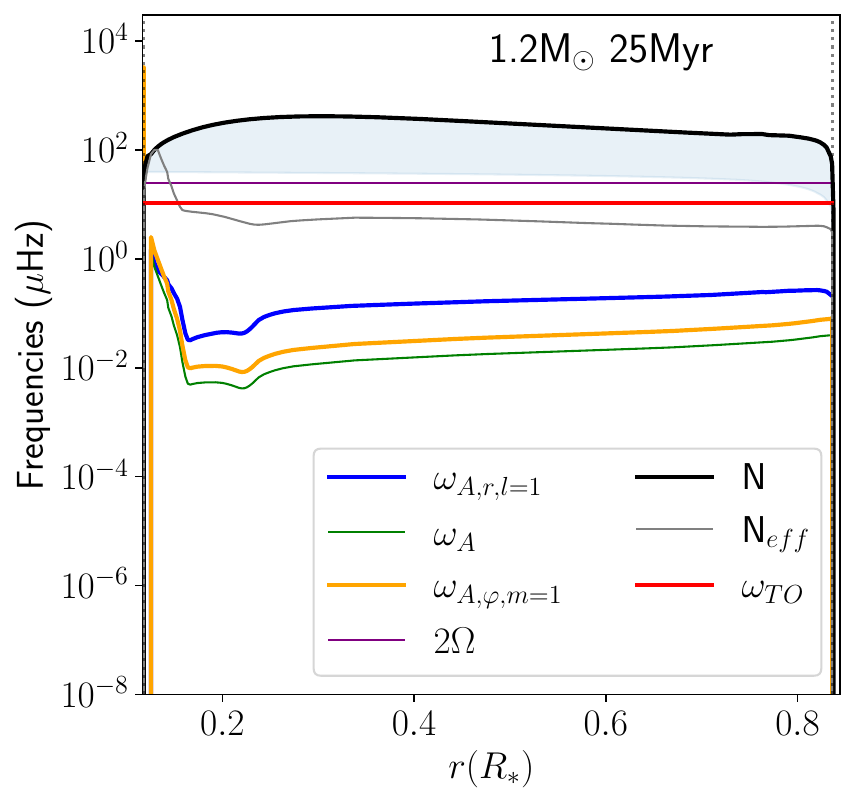}
    \includegraphics[width=0.24\textwidth]{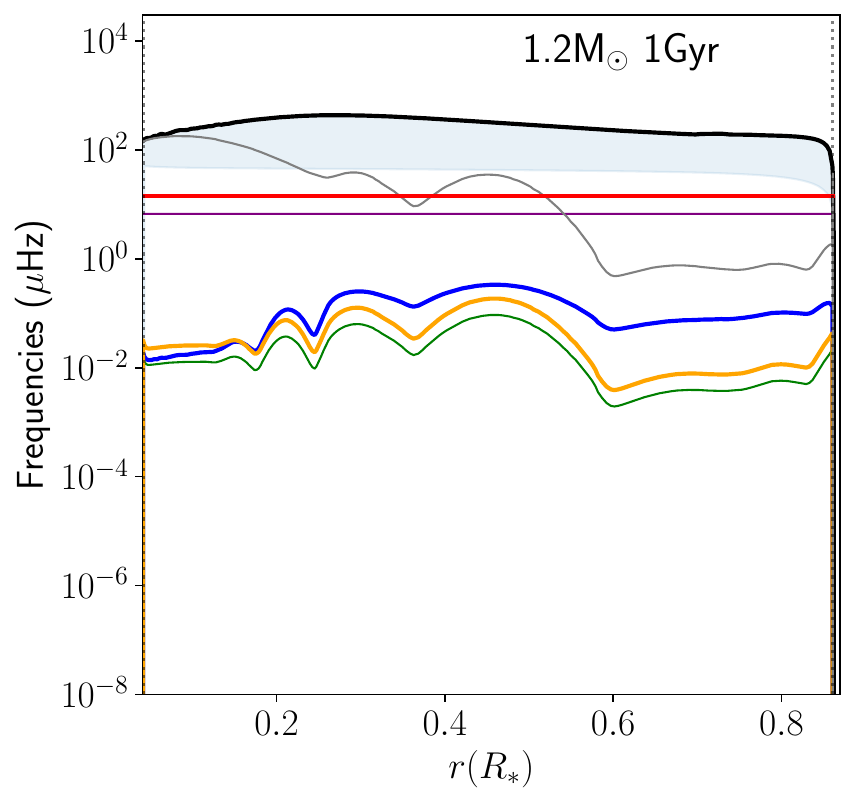}
    \includegraphics[width=0.24\textwidth]{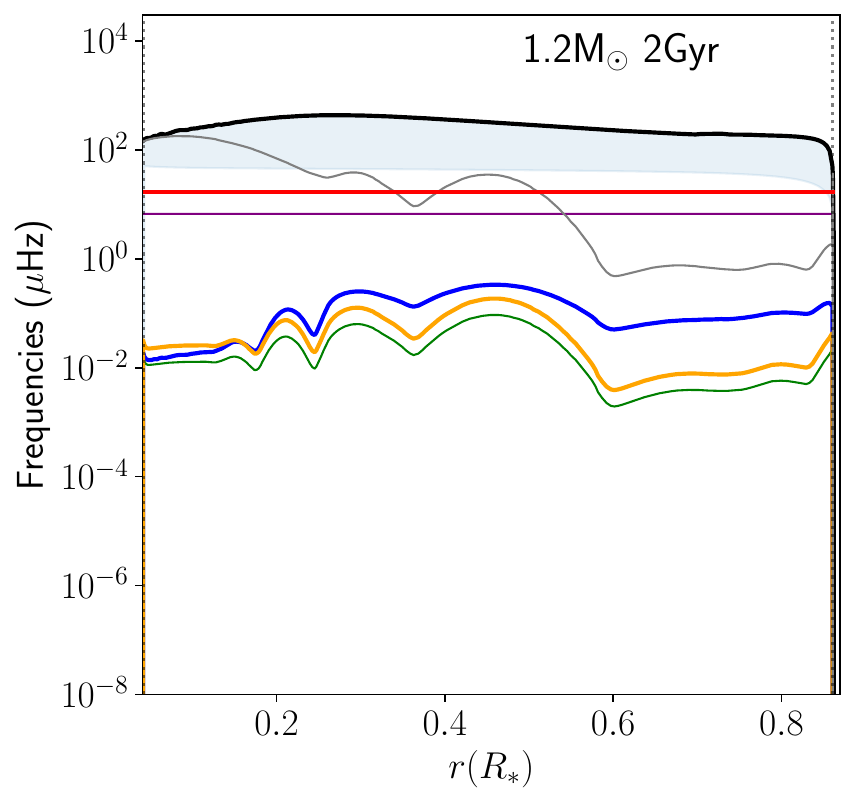}
    \includegraphics[width=0.24\textwidth]{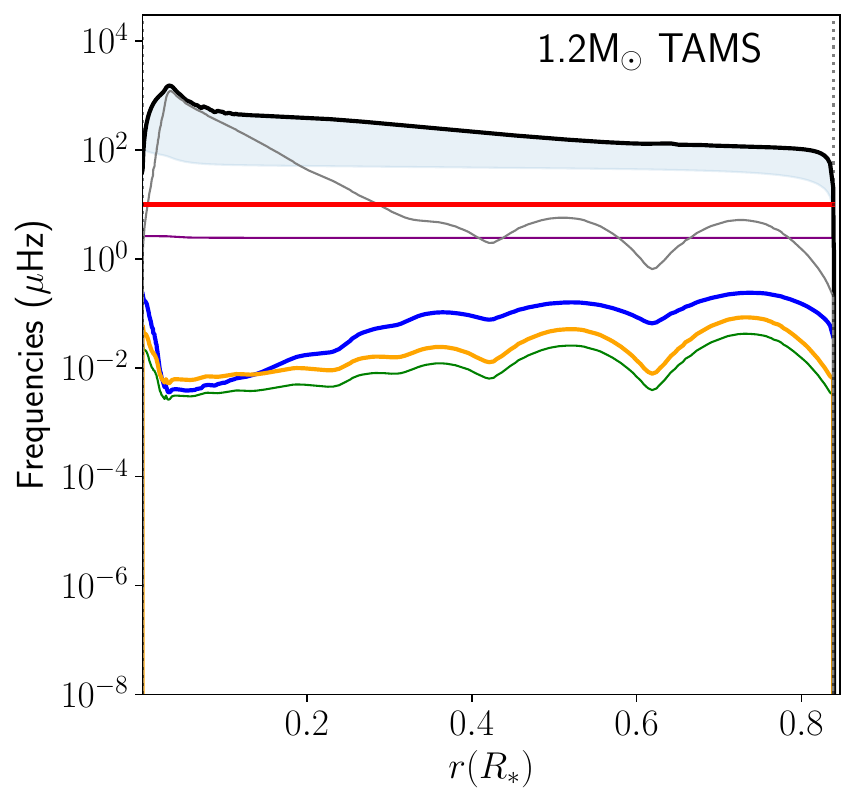}
    \caption{Same as fig.~\ref{fig:allfreq} for a 0.6M$_\odot$ (top line), 0.8M$_\odot$ (middle), and  1.2M$_\odot$ (bottom) star.}
    \label{fig:allfreq060812}
\end{figure*}

On the top and middle panels of figure~\ref{fig:allfreq060812}, we show the various frequencies at 3 different ages for a 0.6 and 0.8 M$_\odot$. 
As expected, since these stars do not reach evolved phases within the universe lifetime and are even more strongly stratified on the main sequence, the effect of a magnetic field generated by a TSD on the propagation of IGW is negligible, as $\omega_{TO} \geq \omega_A$ for the entire evolution. 
Because of their convective envelope going much deeper, the convective turnover frequency of the two models reaches lower value on the main sequence compared to the 1M$_\odot$ case, even below 1$\mu$Hz in the case of the 0.6M$_\odot$ star.
On the ZAMS, the Coriolis frequency is comparable to the effective BV frequency, but the shearing being extremely small, the effective magnetic field generated by TSD stays on the weaker side. 
As the stars evolve, they become more stratified and reach slower rotation rates than the Sun; thus the TSD weakens, and the local Alfv\'en frequency stays way below both the cut-off frequency and the convective turnover frequency until the end of the main sequence. 

\begin{figure*}
    \centering
    \includegraphics[width=0.31\linewidth]{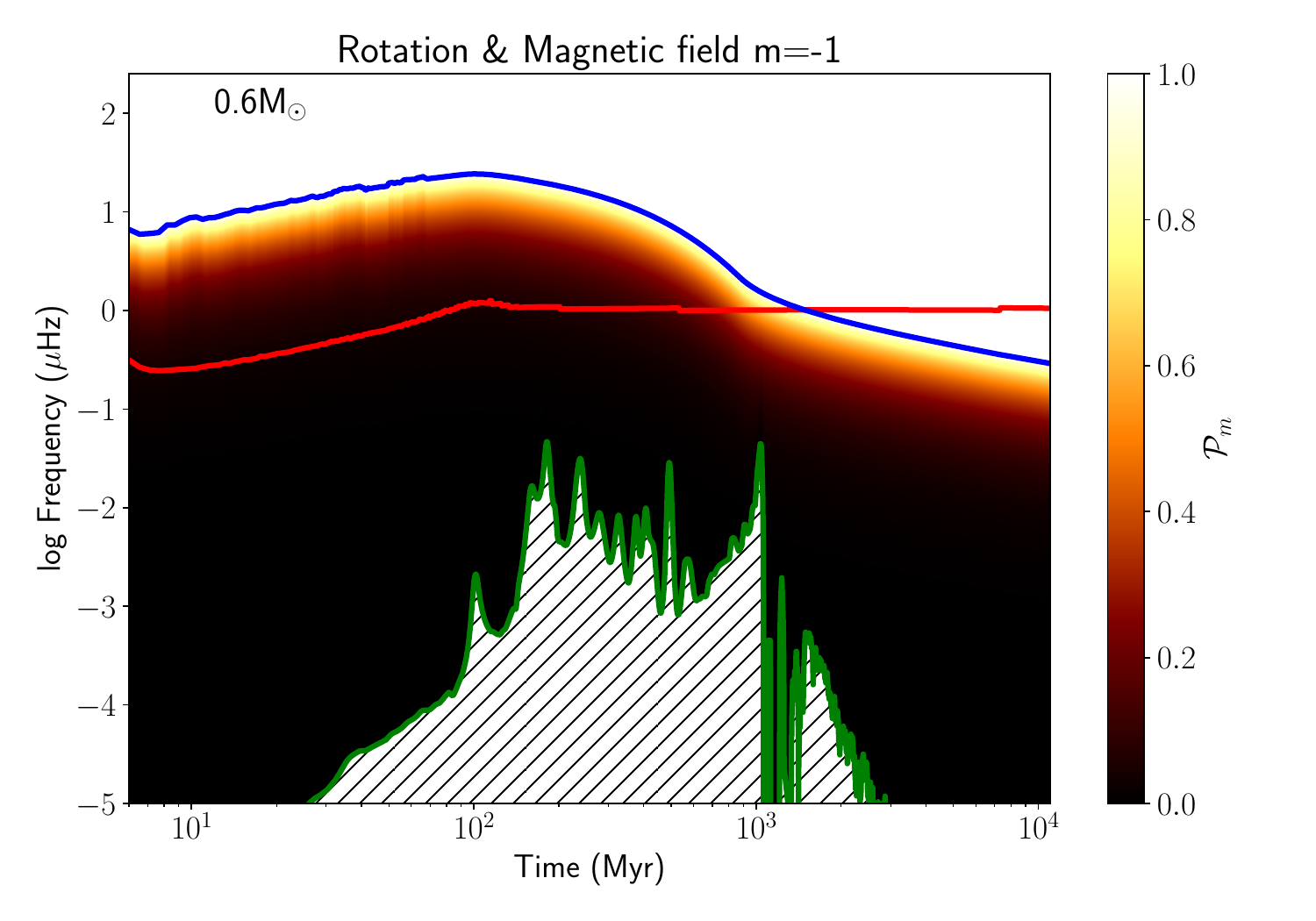}
    \includegraphics[width=0.31\linewidth]{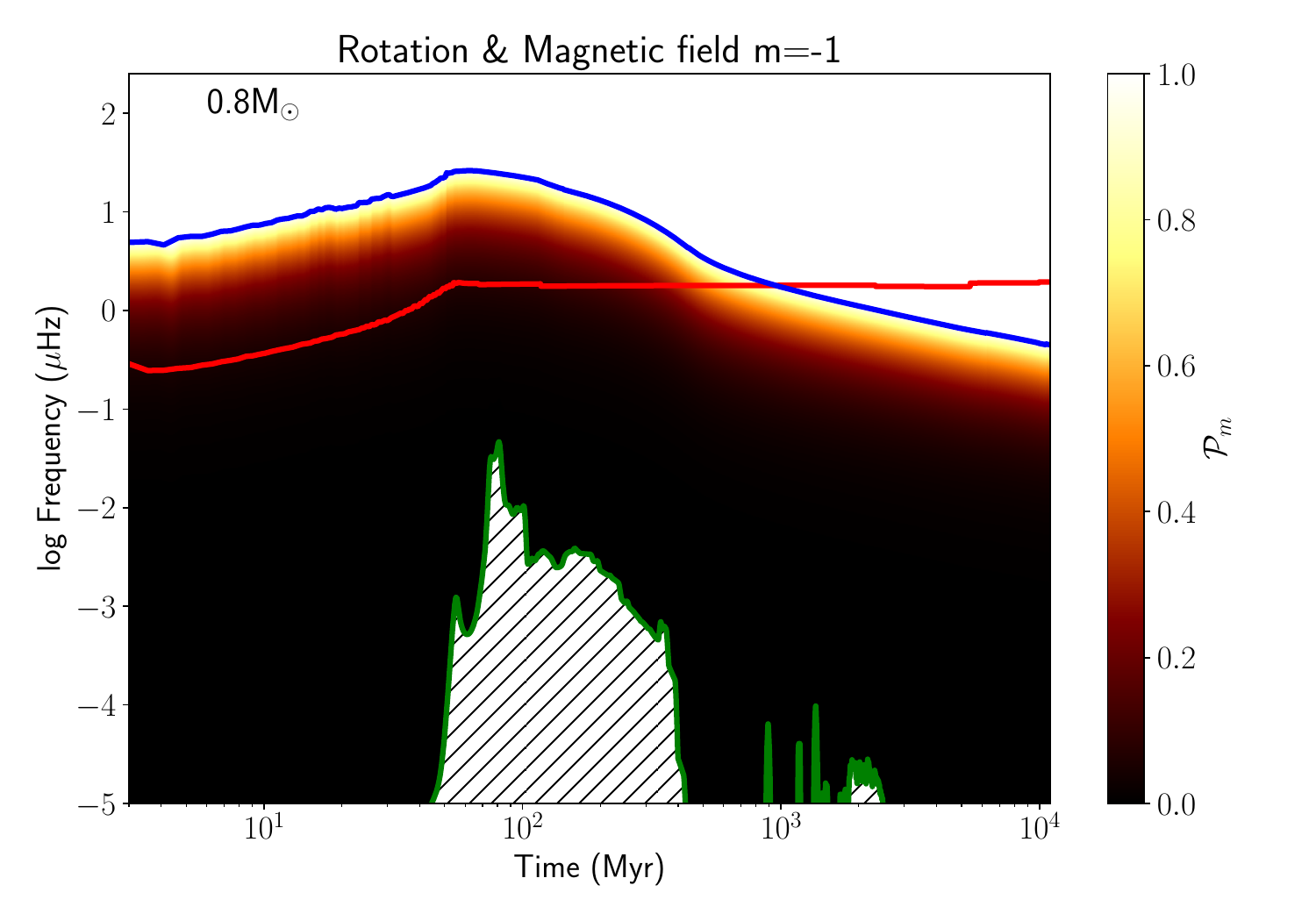}
    \includegraphics[width=0.31\linewidth]{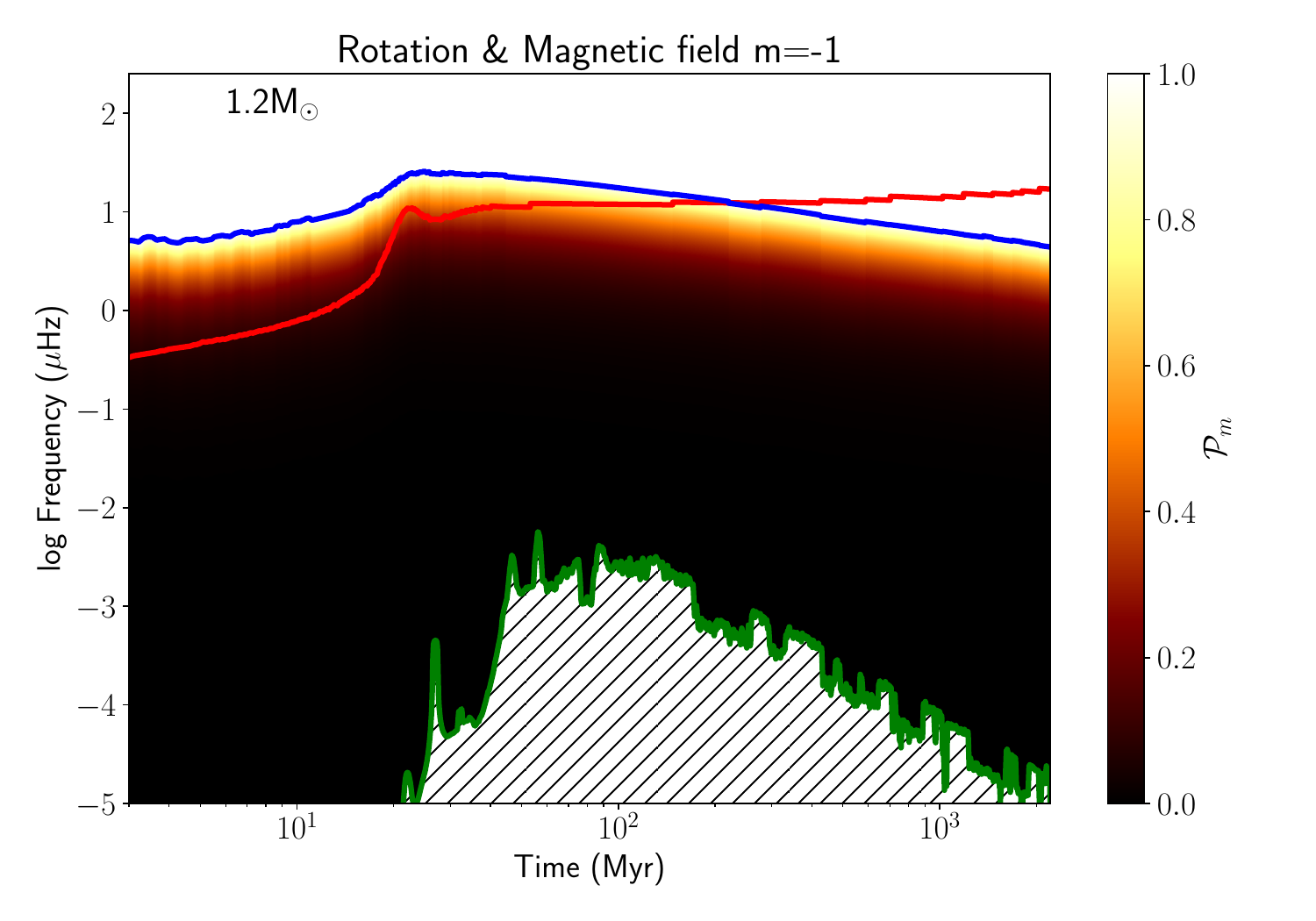}
    \caption{Same as figure~\ref{fig:Pm_1Msun}: from left to right, transmission function $\mathcal{P}_m$ for the 0.6M$_\odot$, 0.8M$_\odot$, and 1.2M$_\odot$ stellar models along their respective evolution, in the case with rotation and magnetism.}
    \label{fig:Pm_LMS}
\end{figure*}

We also explored the case of a 1.2M$_\odot$ star on the main sequence, presented in the bottom line of fig~\ref{fig:allfreq060812}. Although this type of star has a small convective core for most of the main sequence, this does not appear to drastically change the pattern we presented in the case of the 1M$_\odot$ model. 
In this case, the standing gravity modes are not affected by the TSD on the main sequence, since the cutoff frequency is again well above the Alfv\'en frequency. Note that as the star reaches the TAMS, the rotation profile is only mildly differential, confirming the disagreement with asteroseismic observations of the core rotation of subgiant stars that was shown in \cite{Cantiello2014}. 
{Regarding progressive IGWs, since we focused on the main sequence, $\omega_A$ stays well below 1 $\mu$Hz and the convective turnover frequency, showing that a TSD-generated magnetic field has little effect on {most of} the propagation of progressive gravity waves during the main sequence of these stars.
}
We compute the angular momentum transmission function $\mathcal{P}_m$ for the different masses. It is qualitatively similar to the 1M$_\odot$ case but there are some quantitative differences {which are mostly due to the different rotational evolution for a given stellar mass. 
The lower the mass, the longer it takes the star to start spinning down, but the more efficient the spin-down \citep{Mattetal2015}. This is clearly visible on figure~\ref{fig:Pm_LMS} where the 0.6 and 0.8M$_\odot$ stars start to spin down around 100 Myr while the 1.2M$_\odot$ slows down already after 25 Myr, but the formers reach low frequencies, down to 0.5$\mu$Hz at 10 Gyr, while the latter stays relatively fast, a few $\mu$Hz at the end of the MS. In addition, lower-mass stars have a thicker convective envelope and so a longer convective turnover timescale (red line on Fig.~\ref{fig:Pm_LMS}), the bulk of excited waves is thus at lower frequencies. For the 0.6 and the 0.8M$_\odot$ cases, most of the angular momentum flux carried by IGW can be transmitted to the radiative core only after 1 Gyr approximately. Before then, most of the excited flux is clearly sub-inertial and therefore only a very small fraction can be transmitted to the radiative core when compared to the non-rotating non-magnetised case. 
Although the 1.2 M$_\odot$ star is a faster rotator, its small envelope allows for a higher convective turnover frequency. During the PMS, IGW at the turnover frequency do not propagate to the core, but starting at ZAMS, $\mathcal{P}_m$ increases at the turnover frequency and a good part of the convectively-excited flux can cross to the radiative core. Finally, around 300 Myr, most of the IGW flux can be transmitted to the radiative core. \\
We also note that in all cases, the Alfv\'en frequency remains too small to have an impact on the propagation of the excited waves.}

\section{Conclusions}
In this paper, we investigate the role of a magnetic field generated by a Tayler-Spruit Dynamo {following \cite{Spruit2002} formalism} on the propagation of internal gravity waves in the radiative core of {F-, G-, and K-type stars. We computed the evolution of solar-like stars with rotation from the pre-main sequence up to the red-giant branch, including the transport associated to the TSD.}
{While the results highly depend on the prescription we used to model the dynamo, we show that in the case of a TSD modelled according to \citet{Spruit2002}, the produced magnetic field could affect the low-frequency part of the spectrum of progressive gravity waves, in particular as the star goes up the red giant branch, but is too weak to affect {standing} gravity modes. 
The latter point means that there are no direct observable counterparts{. This is because the Alfv\'en frequencies associated to the TSD magnetic field are far inferior to the cut-off frequencies above which the constructive interference of IGW sustains {standing} gravity modes. Therefore, we do not expect the asteroseismic signature to directly capture the TSD, even on the red giant branch where it produces the strongest magnetic field. }
{Nevertheless}, we point out that it may be of greater importance for the transport of angular momentum  during the red giant branch. At this stage, the peak of internal gravity waves excitation falls in the frequency range expected to be converted to magneto-gravity waves, locally affecting the transport and the rotation profile. 
This work presents the first application, on stellar structure models, of a formalism designed to compute a transmission function that characterises the trapping of parts of the internal gravity wave spectrum by rotation and magnetic fields.
We show that the trapping of low-frequency waves is very important during the early evolution, mostly because of the fast rotation rate of young stars that limits the latitudinal extent of the cavity for a large part of the excited spectrum. During later stages of the evolution, the wind braking has spun down the stars enough that most of the excited spectrum becomes super-inertial, largely extending the surface through which the IGW can penetrate in the radiative core to then release their angular momentum. }
The role of the magnetic field at this stage is very limited since it only applies to waves of extremely low frequencies. However, it is expected to play a role -- alongside with rotation -- in the damping of waves in the radiative zone. 
This, however, needs to be explored further through some self-consistent computations including Tayler-Spruit dynamo, transport by internal gravity waves --which become magneto-gravito-inertial waves-- as well as the contribution of the latter to the former.
In a follow-up paper, we will explore the case of intermediate-mass stars such as Gamma Doradus stars for which the rotation profile may be retrieved during the main sequence \citep{VanReeth2016,Ouazzani2017,Christophe2018,Ouazzani2019,Ouazzani2020,Li2020,Moyano2023,Aerts2025} and thus for which additional constraints should be available. 

\begin{acknowledgements}
{We thank the referee, Jim Fuller, for their insightful comments that clearly improved the manuscript.}
L.A. and S.M. acknowledge support from the Centre National des Etudes Spatiales (CNES) through SOHO/GOLF and PLATO grants at CEA/Irfu/DAp-AIM. S.M. acknowledges support from the European Research Council (ERC) under the Horizon Europe program (Synergy Grant agreement 101071505: 4D-STAR) and from PNPS (CNRS/INSU). L.A. acknowledges support from the Swiss National Science Foundation (SNF; Project 200021L-231331) and the French Agence Nationale de la Recherche (ANR-24-CE93-0009-01) ``PRIMA - PRobing the origIns of the Milky WAy’s oldest stars''. While partially funded by the European Union, views and opinions expressed are however those of the authors only and do not necessarily reflect those of the European Union or the European Research Council. Neither the European Union nor the granting authority can be held responsible for them.

\end{acknowledgements}
\bibliography{BibADS}
\appendix
   
\section{Implementation of {the transport of angular momentum associated with} the Tayler-Spruit dynamo in STAREVOL}

\subsection{Corresponding eddy-viscosity and magnetic diffusivity}
\label{Sect:App2}

{We relied on the latest work on the subject and adopted the implementation of the TSD as suggested by \citet{Eggenberger2022} which allows for an easy switch between the formalisms by \citet{Spruit2002} and \cite{Fuller2019}, and provides a calibration of the free parameters attached to the formalism.}

We first determine the Alfvén frequency given here by 
\begin{equation}
    \omega_A = \frac{B}{\sqrt{4\pi\rho}r} .
\end{equation}

To do so, we have to consider that the buoyancy force is weakened by both radiative losses and magnetic diffusivity. The corresponding effective Brunt-V\"ais\"al\"a frequency $N_{\rm eff}$ can be expressed as follows:
\begin{equation}
    N_{\rm eff}^2 = \frac{\eta/K_T}{\eta/K_T + 2}N_T^2 + N_\mu^2, 
    \label{eq:Neff}
\end{equation}
with $\eta$ and $K_T$ the magnetic and thermal diffusivity, respectively. We may also recall the expression of $N_T^2$ and $N_\mu^2$:  
\begin{align}
    N_T^2 = \frac{g\delta}{H_P}(\nabla_{\rm ad}-\nabla),&& 
    N_\mu^2 = \frac{g\varphi}{H_P}\nabla_\mu,
\end{align}
with $g$ and $H_P$ being the local gravity and the pressure height scale, respectively. $\nabla_{\rm ad}$ and $\nabla$ are the adiabatic and local temperature gradients, respectively. Finally, $\delta$ and $\varphi$ are defined as: 
\begin{align}
    \delta = -\left(\frac{\partial \ln \rho}{\partial \ln T}\right)_{P,\mu} &&\textrm{and}&& \varphi = -\left(\frac{\partial \ln \rho}{\partial \ln \mu}\right)_{P,T}.
\end{align}

{The magnetic diffusivity $\eta$ is estimated following the work by \citet{Augustson2019} (see their Eq. 54 in Appendix B) and writes
\begin{equation}
    \eta = 1.02\times 10^{12}\left(2.86Z - 0.97\right)\lambda T^{-3/2},
\end{equation}
where $Z$ is the mean atomic charge, $\lambda$ the Coulomb logarithm is set as 15, and $T$ is the temperature.
To be fully consistent, one should also consider the change in plasma properties associated with the presence of the magnetic instability once it onsets \citep{Spruit2002}.}

{Now, the TSD requires an amplification time $\tau_a$ to be efficient. Spruit (2002) gives 
\begin{equation}
\tau_a = \frac{N_{\rm eff}}{\omega_A\Omega q},
\end{equation}
where 
\begin{equation}
q = \displaystyle\frac{d\ln\Omega}{d\ln r}.
\end{equation}
Based on the magneto-hydrodynamic turbulence formalism introduced by \cite{Fuller2019}, \cite{Eggenberger2022} express the damping timescale of the instability as 
\begin{equation}
    \tau_d = C_\textrm{T}\frac{1}{\omega_A}\left(\frac{\Omega}{\omega_A}\right)^n,
\end{equation}
with $n=1$ to match the prescription by \cite{Spruit2002} and $n=3$ for the expression of \cite{Fuller2019}. Note that these two decay timescales have been qualitatively found in simulations of the TSD by \cite{Petitdemange2023} and by \cite{Barrere2024}, respectively.
The dimensionless number $C_\textrm{T}$ which corresponds to the $\alpha^3$ parameter found in \cite{Fuller2019}'s formalism, has been added to account for uncertainties on the saturated Alfv\'en frequency. In \citet{Eggenberger2022}, they calibrate this parameter by fitting the asteroseismic rotation profile of red giant stars, and we set $C_\textrm{T} = 216$ following their conclusions. 
\\
For the Tayler instability to onset, the shear parameter $q$ needs to be higher than a certain value $q_{\rm min}$. Following \citet{Spruit2002}, we have 
\begin{equation}
    q_{\rm min} = \frac{1}{C_\textrm{T}}\left(\frac{N_{\rm eff}}{\Omega}\right)^{3/2}\left(\frac{\eta}{r^2\Omega}\right)^{1/4}.
\end{equation}}

{We now have $\omega_A$ and $\eta$, and thus $N_{\rm eff}$ thanks to Eq.~(\ref{eq:Neff}). }

{This} allows us to compute the transport of angular momentum associated to the magnetic field. The azimuthal stress {by unit volume} is 
\begin{equation}
    S = \frac{1}{4\pi}B_rB_\phi = \rho r^2 \frac{\omega_A^3}{N_{\rm eff}}.
\end{equation}
The associated viscosity is given by \cite{Spruit2002} {as}
\begin{equation}
    \nu_{\rm TS} = \frac{S}{\rho q\Omega} = \frac{\Omega^2 r^2}{qN_{\rm eff}}\left(\frac{\omega_A}{\Omega}\right)^3.
\end{equation}

We verify in the top panel of figure~\ref{fig:profrot_all} in the main text, showing the rotation profile of the Sun that our implementation {reproduces} the expected rotation profile {constrained by helioseismology}, {as this has been done by} \cite{Eggenberger2005}.

\subsection{Horizontal and vertical Alfv\'en frequencies.}

We {compute} the radial and azimuthal components of the Alfv\'en wave frequency defined as $\omega_A = \Vec{v_A}\cdot\Vec{k}$:
\begin{align}    
    \omega_{A,r}^2 = \frac{B_r^2 k_r^2 }{4\pi\rho} &,&  \omega_{A,\varphi}^2 = \frac{B_\varphi^2 k_\varphi^2}{4\pi\rho}  
\end{align}
with 
\begin{align}    
    B_r^2 = 4\pi\rho \; l_r^2 \; \omega_A^2 &,& B_\varphi^2 = 4\pi\rho \; r^2 \; \omega_A^2, 
\end{align}
where $l_r = r \omega_A/N_{\rm eff}$ and 
\begin{align}    
    k_r^2 = \frac{l(l+1)}{r^2} \left(\frac{N^2}{\omega^2}-1\right) &,& k_\varphi^2 = \frac{m^2}{\sin^2\theta}.
\end{align}
{We maximised the horizontal term by setting $\theta=\pi/2$; it then simplifies }and leads directly to Eq.~(\ref{Eq:omal_comp}).
For the radial term, it comes 
\begin{eqnarray*}
    \omega_{A,r}^2 &=& \frac{B_r^2 k_r^2 }{4\pi\rho} \\
    &=& \frac{4 \pi \rho l_r^2 \omega_A^2}{4 \pi \rho} \left(\frac{N^2}{\omega^2}-1\right)  \frac{l(l+1)}{r^2} \\
    &=& \frac{\omega_A^4}{N_{\rm eff}^2}\left(\frac{N^2}{\omega^2}-1\right) l(l+1). 
\end{eqnarray*}
We then look for the critical frequency where $\omega=\omega_{A,r}$, leading to a very similar result to the transition frequency between the evanescent and propagating regime in the asymptotic regime given in \cite{Fuller2015} :
\begin{equation}
    \omega_{A,r,c}^2 = \omega_{A}^2 \frac{N}{N_{\rm eff}}\sqrt{l(l+1)}.
\end{equation}

\end{document}